\documentclass[a4paper,fleqn,usenatbib]{mnras}

\usepackage[T1]{fontenc}
\usepackage{ae,aecompl}

\usepackage{graphicx}	% Including figure files
\usepackage{amsmath}	% Advanced maths commands
\usepackage{amssymb}	% Extra maths symbols
\usepackage{txfonts}
\usepackage{hyperref}	% Hyperlinks
\hypersetup{colorlinks=true,linkcolor=blue,citecolor=blue,filecolor=blue,urlcolor=blue}

\newcommand{\zerodel}{.\kern-\nulldelimiterspace}

\newcommand{\myvec}{\boldsymbol}

\newcommand{\mytilde}{\raise.17ex\hbox{$\scriptstyle\sim$}}

\newcommand{\deriv}[1]{\frac{\mathrm{d}}{\mathrm{d}#1}}

\newcommand{\oned}{{1\mathrm{D}}}

\newcommand{\twod}{{2\mathrm{D}}}

\newcommand{\threed}{{3\mathrm{D}}}

\newcommand{\numflux}{\myvec{\bar{f}}}

\newcommand{\numfluxcomp}{\bar{f}}

\newcommand{\comment}[1]{}

\title[Discontinuous Galerkin hydrodynamics]{Astrophysical
  hydrodynamics with a high order discontinuous Galerkin scheme and
  adaptive mesh refinement}

\author[K.~Schaal et al.]{\parbox{17cm}{Kevin Schaal$^{1,2}$\thanks{e-mail: kevin.schaal@h-its.org}, Andreas Bauer$^{1}$\protect\thanks{e-mail: andreas.bauer@h-its.org},
  Praveen Chandrashekar$^{3}$, R\"udiger Pakmor$^{1}$, Christian Klingenberg$^4$, Volker Springel$^{1,2}$}
  \vspace*{0.2cm}\\
  $^1$Heidelberg Institute for Theoretical Studies, Schloss-Wolfsbrunnenweg 35, 69118 Heidelberg, Germany\\
  $^2$Zentrum f\"ur Astronomie der Universit\"at Heidelberg,
  Astronomisches Recheninstitut, M\"{o}nchhofstr. 12-14, 69120
  Heidelberg, Germany\\
  $^3$TIFR Centre for Applicable Mathematics, Bangalore-560065, India\\
  $^4$Institut f\"ur Mathematik, Universit\"at W\"urzburg,
  Emil-Fischer-Str. 30, 97074 W\"urzburg, Germany}

\date{Accepted XXX. Received YYY; in original form ZZZ}

\pubyear{2015}

\begin{document}
\label{firstpage}
\pagerange{\pageref{firstpage}--\pageref{lastpage}}
\maketitle

\begin{abstract}
  Solving the Euler equations of ideal hydrodynamics as accurately and
  efficiently as possible is a key requirement in many astrophysical
  simulations. It is therefore important to continuously advance the
  numerical methods implemented in current astrophysical codes,
  especially also in light of evolving computer technology, which
  favours certain computational approaches over others. Here we
  introduce the new adaptive mesh refinement (AMR) code {\small
    TENET}, which employs a high order discontinuous Galerkin (DG)
  scheme for hydrodynamics.  The Euler equations in this method are
  solved in a weak formulation with a polynomial basis by means of
  explicit Runge-Kutta time integration and Gauss-Legendre
  quadrature. This approach offers significant advantages over
  commonly employed second order finite volume (FV) solvers. In particular, the
  higher order capability renders it computationally more efficient,
  in the sense that the same precision can be obtained at
  significantly less computational cost. Also, the DG scheme
  inherently conserves angular momentum in regions where no limiting
  takes place, and it typically produces much smaller numerical
  diffusion and advection errors than a FV approach.  A further
  advantage lies in a more natural handling of AMR refinement
  boundaries, where a fall-back to first order can be
  avoided. Finally, DG requires no wide stencils at high order, and
  offers an improved data locality and a focus on local computations, 
  which is favourable for current and upcoming highly parallel
  supercomputers. We describe the formulation and implementation
  details of our new code, and demonstrate its performance and
  accuracy with a set of two- and three-dimensional test problems. The
  results confirm that DG schemes have a high potential for
  astrophysical applications.
\end{abstract}


\section{Introduction}
\label{sec:introduction}

Through the availability of ever more powerful computing resources,
computational fluid dynamics has become an important part of
astrophysical research. It is of crucial help in shedding light on the
physics of the baryonic part of our Universe, and contributes
substantially to progress in our theoretical understanding of galaxy
formation and evolution, of star and planet formation, and of
the dynamics of the intergalactic medium. In addition, it plays an
important role in planetary science and in solving engineering
problems related to experimental space exploration.

In the astrophysics community, most numerical work thus far on solving
the Euler equations of ideal hydrodynamics has been carried out with
two basic types of codes. On the one hand, there is the broad class of
Eulerian methods which utilize classical hydrodynamic solvers
operating on fixed Cartesian grids \citep[e.g.][]{ATHENA}, or on
meshes which can adjust their resolution in space with the adaptive
mesh refinement (AMR) technique \citep[e.g.][]{FLASH, RAMSES, PLUTO,
  ENZO}. On the other hand, there are pseudo-Lagrangian discretizations in the
form of smoothed particle hydrodynamics (SPH), which are other flexible and 
popular tools to study many astrophysical problems
\citep[e.g.][]{GASOLINE, GADGET2}.

Some of the main advantages and drawbacks of these methods become
apparent if we recall the fundamental difference in their numerical
approach, which is that grid codes discretize space whereas SPH
decomposes a fluid in terms of mass elements. The traditional
discretization of space used by Eulerian methods yields good
convergence properties, and a high accuracy and efficiency for many
problems.  Furthermore, calculations can be straightforwardly
distributed onto parallel computing systems, often allowing a high
scalability provided there is only little communication between the
cells.  The discretization of mass used by SPH on the other hand
results in a natural resolution adjustment in converging flows such
that most of the computing time and available resolution are dedicated
to dense, astrophysically interesting regions.  Moreover, Lagrangian
methods can handle gas with high advection velocities without
suffering from large errors.  However, both methods have also
substantial weaknesses, ranging from problems with the
Galilean invariance of solutions in the case of grid codes, to a
suppression of fluid instabilities and noise in the case of SPH.

This has motivated attempts to combine the advantages of traditional
grid-based schemes and of SPH in new methods, such as moving
mesh-codes \citep{AREPO, TESS} or in mesh-free methods that retain a
higher degree of accuracy \citep{LANSON2008, GIZMO} than SPH.  Both of these new
developments conserve angular momentum better than plain Eulerian
schemes, while still capturing shocks accurately, a feature that is
crucial for many applications in astrophysics. However, these new
methods need to give up the simple discretization of space into
regular grids, meaning also that their computational efficiency takes
a significant hit, because the regular memory access patterns possible
for simple structured discretizations are ideal for leveraging a high
fraction of the peak performance of current and future hardware.

In fact, the performance of supercomputers has increased exponentially
over the last two decades, roughly following the empirical trend of
Moore's law, which states that the transistor count and hence
performance of computing chips doubles roughly every two years.  Soon,
parallel computing systems featuring more than 10 million cores and
``exascale'' performance in the range of $10^{18}$ floating point
operations per second are expected. Making full use of this enormous
compute power for astrophysical research will require novel
generations of simulation codes with superior parallel scaling and
high resilience when executed on more and more cores. Also, since the
raw floating point speed of computer hardware grows much faster than
memory access speed, it is imperative to search for new numerical
schemes 
that focus on local computations within the resolution elements.
Those allow for data locality in the memory leading to a fast data transfer to the 
CPU cache, and furthermore, the relative cost for 
communication with neighbouring resolution elements is reduced.

A very interesting class of numerical methods in this context are
so-called discontinuous Galerkin (DG) schemes, which can be used for a
broad range of partial differential equations.  Since the introduction
of DG \citep{REED1973} and its generalization to nonlinear problems
\citep{CS2, CS1, CS5, CS3, CS4}, it has been successfully applied in
diverse fields of physics such as aeroacoustics, electro magnetism,
fluid dynamics, porous media, etc.~\citep{DGBOOK, GALLEGO2014}. 
On the other hand, in the astrophysics community the
adoption of modern DG methods has been fairly limited so far.
However, two recent works suggest that this is about to
change. \citet{MOCZ2014} presented a DG method for solving the
magnetohydrodynamic (MHD) equations on arbitrary grids as well as on a
moving Voronoi mesh, and \citet{ZANOTTI2015} developed an AMR code for
relativistic MHD calculations. The present paper is a further
contribution in this direction and aims to introduce a novel DG-based
hydrodynamical code as an alternative to commonly employed schemes in
the field.

DG is a finite element method which incorporates several aspects from
finite volume (FV) methods. The partial differential equation is
solved in a weak formulation by means of local basis functions,
yielding a global solution that is in general discontinuous across
cell interfaces.  The approach requires communication only between
directly neighbouring cells and allows for the exact conservation of
physical quantities including angular momentum. Importantly, the
method can be straightforwardly implemented with arbitrary spatial
order, since it directly solves also for higher order moments of the
solution. Unlike in standard FV schemes, this higher order accuracy is
achieved without requiring large spatial stencils, making DG
particularly suitable for utilizing massive parallel systems with
distributed memory because of its favourable compute-to-communicate
ratio and enhanced opportunities to hide communication behind local
computations.

In order to thoroughly explore the utility of DG in real
astrophysical applications, we have developed {\small TENET}, an
MPI-parallel DG code which solves the Euler equations on an AMR grid
to arbitrary spatial order. In our method, the solution within every
cell is given by a linear combination of Legendre polynomials, and the
propagation in time is accomplished with an explicit Runge-Kutta (RK)
time integrator.  A volume integral and a surface integral have to be computed
numerically for every cell in every timestep. The surface integral
involves a numerical flux computation which we carry out with a
Riemann solver, similar to how this is done in standard Godunov
methods. In order to cope with physical discontinuities and spurious
oscillations we use a simple minmod limiting scheme. 

The goal of this paper is to introduce the concepts of our DG
implementation and compare its performance to a standard FV method
based on the reconstruct-solve-average (RSA) approach.  In this
traditional scheme, higher order information is discarded in the
averaging process and recomputed in the reconstruction step, leading
to averaging errors and numerical diffusion. DG on the other hand does
not only update the cell-averaged solution every cell, but also higher order
moments of the solution, such that the reconstruction becomes
obsolete. This aspect of DG leads to important advantages over FV
methods, in particular an inherent improvement of angular momentum
conservation and much reduced advection errors, especially for but not
restricted to smooth parts of the solutions. These accuracy gains and
the prospect to translate them to a lower computational cost at given
computational error form a strong motivation to investigate the use of
DG in astrophysical applications.

The paper at hand is structured as follows. In
Section~\ref{sec:dg_hydrodynamics}, we present the general methodology
of our DG implementation for Cartesian grids. The techniques we
adopt for limiting the numerical solution are described in
Section~\ref{sec:slope_limiting}, and the generalization to a mesh
with adaptive refinement is outlined in Section~\ref{sec:dg_amr}.
These sections give a detailed account of the required equations and
discretization formulae for the sake of clarity and definiteness,
something that we hope does not discourage interested readers. We then
validate our DG implementation and compare it to a standard second order Godunov FV
solver with two- and three-dimensional test problems in
Section~\ref{sec:validation}. Finally, Section \ref{sec:summary}
summarizes our findings and gives a concluding discussion.

\section{Discontinuous Galerkin hydrodynamics}
\label{sec:dg_hydrodynamics}

\subsection{Euler equations}

The Euler equations are conservation laws for mass, momentum, and
total energy of a fluid. They are a system of hyperbolic
partial differential equations and can be written in compact form as
\begin{align}
\frac{\partial\myvec{u}}{\partial{t}}+\sum\limits_{\alpha=1}^{3}\frac{\partial\myvec{f}_\alpha}{\partial x_\alpha}=0,
%\frac{\partial}{\partial{t}}\myvec{u(\myvec{x},t)}+\sum\limits_{\alpha=1}^{3}\frac{\partial}{\partial x_\alpha}\myvec{f}_\alpha(\myvec{u}(\myvec{x},t))=0,
\label{eq:euler}
\end{align}
with the state vector 
\begin{align}
\myvec{u}=\begin{pmatrix}\rho \\ \rho\myvec{v} \\ \rho e \end{pmatrix}=\begin{pmatrix}\rho \\ \rho\myvec{v} \\ \rho u + \frac{1}{2}\rho\myvec{v}^2 \end{pmatrix},
\label{eq:state_vector}
\end{align}
and the flux vectors
\begin{align}
\myvec{f}_1=\begin{pmatrix} \rho v_1 \\ \rho v_1^2+p \\ \rho v_1 v_2 \\ \rho v_1 v_3 \\ (\rho e + p) v_1\end{pmatrix} \,\,\,\,
\myvec{f}_2=\begin{pmatrix} \rho v_2 \\ \rho v_1 v_2 \\ \rho v_2^2+p\\ \rho v_2 v_3 \\ (\rho e + p) v_2\end{pmatrix} \,\,\,\,
\myvec{f}_3=\begin{pmatrix} \rho v_3 \\ \rho v_1 v_3\\ \rho v_2 v_3 \\ \rho v_3^2+p \\ (\rho e + p)v_3\end{pmatrix}.
\label{eq:flux_vectors}
\end{align}
%\myvec{u}=\myvec{u}\left(\myvec{x},t\right)$
The unknown quantities are density $\rho$,
velocity $\myvec{v}$,
pressure $p$,
and total energy per unit mass $e$.
The latter can be expressed in terms of the internal energy per unit
mass $u$
and the kinetic energy of the fluid, $e=u+\frac{1}{2}\myvec{v}^2$.  For an ideal gas, the system is
closed with the equation of state
\begin{align}
p=\rho u (\gamma-1),
\label{eq:eos}
\end{align}
where $\gamma$ denotes the adiabatic index. 

\subsection{Solution representation} 

We partition the domain by non-overlapping
cubical cells, which may be refined using AMR techniques
as explained in later sections.
Moreover, we follow the approach of a classical modal DG scheme, where
the solution in cell $K$ is given by a linear combination of $N(k)$ orthogonal 
and normalized basis functions $\phi_l^K$:
\begin{align}
\myvec{u}^K\left(\myvec{x},t\right)=\sum\limits_{l=1}^{N(k)}\myvec{w}_l^K(t)\phi_l^K(\myvec{x}).
\label{eq:solution_representation}
\end{align}
In this way, the dependence on time and space of the solution is split
into time-dependent weights, and basis functions which are constant in
time.  Consequently, the state of a cell is completely characterized
by $N(k)$ weight vectors $\myvec{w}_j^K(t)$.

The above equation can be solved for the weights by multiplying with
the corresponding basis function $\phi_j^K$ and integrating over the
cell volume. Using the orthogonality and normalization of the basis
functions yields
\begin{align}
\myvec{w}^K_j=\frac{1}{|K|}\int_K\!\myvec{u}^K\phi^K_j\,\mathrm{d}V,\quad j=1,\ldots,N(k),
\label{eq:weights}
\end{align}
where $|K|$ is the volume of the cell.  The first basis function is
chosen to be $\phi_1=1$ and hence the weight $\myvec{w}_1^K$ is the
cell average of the state vector $\myvec{u}^K$. The higher order
moments of the state vector are described by weights $\myvec{w}_j^K$
with $j\ge2$.  

The basis functions can be defined on a cube in terms of scaled
variables $\myvec{\xi}$,
\begin{align}
\phi_l(\myvec{\xi}):\left[-1,1\right]^3\rightarrow\mathbb{R}.
\end{align}
The transformation between coordinates $\myvec{\xi}$ in the cell frame of reference
and coordinates $\myvec{x}$ in the laboratory frame of reference is
\begin{align}
\myvec{\xi}=\frac{2}{\Delta x^K}(\myvec{x}-\myvec{x}^K),
\label{eq:trafo}
\end{align}
where $\Delta x^K$ and $\myvec{x}^K$ are edge length and cell centre of cell $K$, respectively.
For our DG implementation, we construct a set of three-dimensional polynomial basis functions with a maximum degree of $k$ as products of one-dimensional
scaled Legendre polynomials $\tilde{P}$:
\begin{align}
\left\{\phi_l(\myvec{\xi})\right\}_{l=1}^{N(k)}=\left\{\tilde{P}_u(\xi_1)\tilde{P}_v(\xi_2)\tilde{P}_w(\xi_3)|u,v,w\in\mathbb{N}_0\wedge u+v+w\le k\right\}.
\label{eq:set_basis_functions}
\end{align}
The first few Legendre polynomials are shown in Appendix \ref{sec:legendre_polynomials}.
The number of basis functions for polynomials with a maximum degree of $k$ is
\begin{align}
N(k)=\sum\limits_{u=0}^k\sum\limits_{v=0}^{k-u}\sum\limits_{w=0}^{k-u-v}1=\frac{1}{6}(k+1)(k+2)(k+3).
\label{eq:nof_basis_functions}
\end{align}
Furthermore, when polynomials with a maximum degree of $k$ are used, a
scheme with spatial order $p=k+1$ is obtained. For example, linear
basis functions lead to a scheme which is of second order in space.

\subsection{Initial conditions}

Given initial conditions $\myvec{u}(\myvec{x},t=0)=\myvec{u}(\myvec{x},0)$,
we have to provide an initial state for the DG scheme which is
consistent with the solution representation.
To this end, the initial conditions are expressed by means of
the polynomial basis on cell $K$, which will then be
\begin{align}
\myvec{u}^K\left(\myvec{x},0\right)=\sum\limits_{l=1}^{N(k)}\myvec{w}_l^K(0)\phi_l^K(\myvec{x}).
\label{eq:initial_solution_representation}
\end{align}
If the initial conditions at hand are
polynomials with degree $\le k$, this representation
preserves the exact initial conditions,
otherwise equation~\eqref{eq:initial_solution_representation}  is an approximation to the given initial conditions.
The initial weights can be obtained by performing an $L^2$-projection,
\begin{align}
\min_{\left\{w^K_{l,i}(0)\right\}_l}\int_{K}\!\left(u_i^K(\myvec{x},0)-u_i(\myvec{x},0)\right)^2\,\mathrm{d}V,\quad i=1,\ldots,5,
\end{align}
where $i=1,\ldots,5$ enumerates the conserved variables.  The projection above leads to the integral
\begin{align}
\myvec{w}^K_j(0)=\frac{1}{|K|}\int_K\!\myvec{u}(\myvec{x},0)\phi^K_j(\myvec{x})\,\mathrm{d}V,\quad j=1,\ldots,N(k),
\end{align}
which can be transformed to the reference frame of the cell, viz.
\begin{align}
\myvec{w}^K_j(0)=\frac{1}{8}\int_{[-1,1]^3}\!\myvec{u}(\myvec{\xi},0)\phi_j(\myvec{\xi})\,\mathrm{d}\myvec{\xi},\quad j=1,\ldots,N(k).
\end{align}
We solve the integral numerically by means of tensor product Gauss-Legendre quadrature (hereafter called Gaussian quadrature) with $(k+1)^3$ nodes:
\begin{align}
\myvec{w}^K_j(0)\approx\frac{1}{8}\sum_{q=1}^{(k+1)^3}\myvec{u}(\myvec{\xi}_q^{\threed},0)\phi_j(\myvec{\xi}_q^{\threed})\omega_q^{\threed},\quad j=1,\ldots,N(k).
\end{align}
Here, $\myvec{\xi}_q^{\threed}$ is the position of
the quadrature node $q$ in the cell frame of reference
and $\omega_q^{\threed}$ denotes the
corresponding quadrature weight.  The technique of Gaussian quadrature is
explained in more detail in Appendix~\ref{sec:gaussian_quadrature}.
\begin{figure}
\centering
\includegraphics{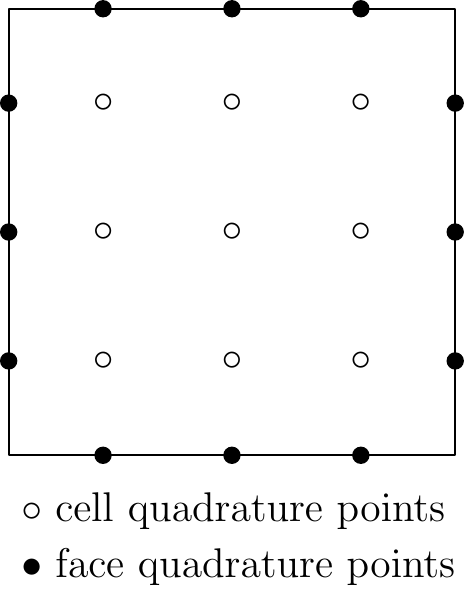}
\caption{In the DG scheme, a surface integral and a volume integral have to be
  computed numerically for every cell, see
  equation~\eqref{eq:dg_projected}.  In our approach, we solve these
  integrals by means of Gaussian quadrature
  (Appendix~\ref{sec:gaussian_quadrature}). This example shows the
  nodes for our third order DG method (with second order polynomials)
  when used in a two-dimensional configuration.  The black nodes
  indicate the positions where the surface integral is evaluated,
  which involves a numerical flux calculation with a Riemann solver.
  The white nodes are used for numerically estimating the volume
  integral. }
\label{fig:gaussquad}
\end{figure}

\subsection{Evolution equation for the weights} 

In order to derive the DG scheme on a cell $K$, the Euler equations
for a polynomial state vector $\myvec{u}^K$
are multiplied by the basis function $\phi^K_j$ and integrated over the
cell volume,
\begin{align}
\int_K\!\left[\frac{\partial\myvec{u}^K}{\partial{t}}+\sum\limits_{\alpha=1}^{3}\frac{\partial\myvec{f}_\alpha}{\partial x_\alpha}\right]\phi_j^K\,\mathrm{d}V=0.
\label{eq:euler_project}
\end{align}
Integration by parts of the flux divergence term and a subsequent application of Gauss' theorem leads to
\begin{align}
\frac{\mathrm{d}}{\mathrm{d}{t}}\int_K\!\myvec{u}^K\phi_j^K\,\mathrm{d}V-\sum_{\alpha=1}^3\int_K\!\myvec{f}_\alpha\frac{\partial \phi_j^K}{\partial x_\alpha}\,\mathrm{d}V
+\sum_{\alpha=1}^3\int_{\partial K}\myvec{f}_\alpha n_{\alpha}\phi_j^K\,\mathrm{d}S=0,
\label{eq:dg_projected}
\end{align}
where $\hat{n}=(n_1,n_2,n_3)^T$ denotes the outward-pointing unit normal vector of the surface $\partial K$.
In the following, we discuss each of the terms separately.

According to equation~\eqref{eq:weights}, the first term is simply the
time variation of the weights,
\begin{align}
\frac{\mathrm{d}}{\mathrm{d}t}\int_K\!\myvec{u}^K\phi_j^K\,\mathrm{d}V=|K|\frac{\mathrm{d}\myvec{w}_j^K}{\mathrm{d}t}.
\label{eq:dg_term1}
\end{align}
The second and third terms are discretized by transforming the
integrals to the cell frame and applying Gaussian quadrature
(Fig.~\ref{fig:gaussquad}, Appendix \ref{sec:gaussian_quadrature}).
With this procedure the second term becomes
\begin{align}
\nonumber&\sum_{\alpha=1}^3\int_K\!\myvec{f}_\alpha\left(\myvec{u}^K(\myvec{x},t)\right)\frac{\partial \phi_j^K(\myvec{x})}{\partial x_\alpha}\,\mathrm{d}V\\
\nonumber&=\frac{(\Delta x^K)^2}{4}\sum_{\alpha=1}^3\int_{[-1,1]^3}\!\myvec{f}_\alpha\left(\myvec{u}^K(\myvec{\xi},t)\right)\frac{\partial \phi_j(\myvec{\xi})}{\partial \xi_\alpha}\,\mathrm{d}\myvec{\xi}\\
&\approx\frac{(\Delta x^K)^2}{4}\sum_{\alpha=1}^3\sum_{q=1}^{(k+1)^3}\myvec{f}_\alpha\left(\myvec{u}^K(\myvec{\xi}_q^{\threed},t)\right)\left\zerodel\frac{\partial \phi_j(\myvec{\xi})}{\partial \xi_\alpha}\right|_{\myvec{\xi}_q^{\threed}}\omega_q^{\threed}.
\label{eq:dg_term2}
\end{align}
Note that the transformation of the derivative
$\partial/\partial x_\alpha$ gives a factor of $2$, see
equation~\eqref{eq:trafo}.  The volume integral is computed by
Gaussian quadrature with $k+1$ nodes per dimension. These nodes allow
the exact integration of polynomials up to degree $2k+1$.

The flux functions $\myvec{f}_\alpha$ in the above expression can be evaluated
analytically, which is not the case for the fluxes in the last term of
the evolution equation~\eqref{eq:dg_projected}. This is because the
solution is discontinuous across cell interfaces. We hence have to
introduce a numerical flux function
$\numflux\left(\myvec{u}^{K-},\myvec{u}^{K+},\hat{n}\right)$, which in
general depends on the states left and right of the interface and on
the normal vector. With this numerical flux, the third term in
equation~\eqref{eq:dg_projected} takes the form
\begin{align}
\nonumber &\sum_{\alpha=1}^3\int_{\partial K}\myvec{f}_\alpha n_{\alpha}(\myvec{x})\phi_j^K(\myvec{x})\,\mathrm{d}S\\
\nonumber &=\frac{(\Delta x^K)^2}{4}\int_{\partial [-1,1]^3}\numflux\left(\myvec{u}^{K-}(\myvec{\xi},t),\myvec{u}^{K+}(\myvec{\xi},t),\hat{n}(\myvec{\xi})\right)\phi_j(\myvec{\xi})\,\mathrm{d}S_{\myvec{\xi}}\\
&\approx\frac{(\Delta x^K)^2}{4}\sum_{A\in \partial [-1,1]^3}\sum_{q=1}^{(k+1)^2}\numflux\left(\myvec{u}^{K-}(\myvec{\xi}_{q,A}^{\twod},t),\myvec{u}^{K+}(\myvec{\xi}_{q,A}^{\twod},t),\hat{n}\right)\phi_j(\myvec{\xi}_{q,A}^{\twod})\omega_q^{\twod}.
\label{eq:dg_term3}
\end{align}
Here for each interface of the normalized cell, a two-dimensional
Gaussian quadrature rule with $(k+1)^2$ nodes is applied.  The
numerical flux across each node can be calculated with a
one-dimensional Riemann solver, as in ordinary Godunov schemes. For
our DG scheme, we use the positivity preserving HLLC Riemann solver
\citep{TORO}.

In order to model physics extending beyond ideal
hydrodynamics, source terms can be added on the right-hand side
of equation~\eqref{eq:euler_project}. Most importantly, the treatment
of gravity is accomplished by the source term
\begin{align}
\myvec{s}=\begin{pmatrix} 0 \\-\rho\myvec{\nabla}\Phi \\ -\rho\myvec{v}\cdot\myvec{\nabla}\Phi\end{pmatrix},
\label{eq:gravity_term}
\end{align}
which by projecting onto the basis function $\phi_j$ inside cell $K$ and discretizing 
becomes
\begin{align}
\nonumber&\int_K\!\myvec{s}(\myvec{x},t)\phi_j^K({\myvec{x}})\,\mathrm{d}V\\
\nonumber&=\frac{|K|}{8}\int_{[-1,1]^3}\!\myvec{s}(\myvec{\xi},t)\phi_j(\myvec{\xi})\,\mathrm{d}\myvec{\xi}\\
&\approx\frac{|K|}{8}\sum_{q=1}^{(k+1)^3}\myvec{s}(\myvec{\xi}_q^{\threed},t)\phi_j(\myvec{\xi}_q^{\threed})\omega_q^{\threed}.
\end{align}

We have now discussed each term of the basic
equation~(\ref{eq:dg_projected}) and arrived at a spatial
discretization of the Euler equations of the form
\begin{align}
\frac{\mathrm{d}\myvec{w}_j^K}{\mathrm{d}t}+\myvec{R}_K=0,\quad j=1,\ldots,N(k),
\label{eq:spatial_discretisation}
\end{align}
which represents a system of coupled ordinary differential equations.
For the discretization in time we apply an explicit strong stability
preserving Runge-Kutta (SSP RK) scheme \citep{GOTTLIEB2001} 
of the same order as the spatial discretization. 
With a combination of an SSP RK method, a
positivity preserving Riemann solver, and a positivity limiter (see
Section~\ref{sec:positivity_limiter}), negative pressure and density values in
the hydro scheme can be avoided. The Butcher tables of the first- to
fourth order SSP RK methods we use in our code are listed in
Appendix~\ref{sec:runge_kutta}.

\subsection{Timestep calculation}

If the Euler equations are solved with an 
explicit time integrator, the timestep has to fulfil a
Courant-Friedrichs-Lewy (CFL) condition for achieving numerical timestep
stability. For the DG scheme with explicit RK time integration, 
the timestep depends also on the order $p=k+1$ of the scheme. 
We calculate the timestep $\Delta t^K$ of cell $K$ following \citet{CS2} as
\begin{align}
\Delta t^K=\frac{\text{CFL}}{2k+1}\left(\frac{|v_1^K|+c^K}{\Delta x_1^K}+\frac{|v_2^K|+c^K}{\Delta x_2^K}+\frac{|v_3^K|+c^K}{\Delta x_3^K}\right)^{-1},
\label{eq:time_step}
\end{align}
where $c=\sqrt{\gamma p / \rho}$ is the sound speed and the
denominators in the parantheses are the edge lengths of the cell.  Since
we used only cubic cells, we have
$\Delta x_1^K = \Delta x_2^K = \Delta x_3^K = \Delta x^K$.  In this
work, we apply a global timestep given by the minimum of the local
timesteps among all cells. The CFL number depends on the choice of
flux calculation but in practice should be set to a somewhat smaller
value as formally required. 
This is because 
for the calculation of the timestep in an RK method of 
second order or higher, a mathematical inconsistency arises. 
At the beginning of the timestep, only the current
velocity and sound speed of the gas are known, whereas in principle the
maximum speed occurring during all RK stages within one step should be used 
for calculating the CFL timestep to be strictly safe. 
This information is not readily
available, and the only straightforward way to cope with this problem
is to reduce the CFL number to a value smaller than mathematically required. 
For the tests presented in this paper,
we decided to adopt a
conservative choice of $\text{CFL}=0.2$.

Furthermore, if the positivity limiter is used, the hydro timestep
$\Delta t^K$ has to be modified and can be slightly more restrictive,
as described in Section~\ref{sec:positivity_limiter}.

Source terms $\myvec{s}(\myvec{u}, t)$ on the right hand side of the
Euler equations can induce additional timestep criteria.  In this
case, positivity of the solution can be enforced by determining the
timestep such that $\rho(\myvec{u}')>0$ and $p(\myvec{u}')>0$, with
$\myvec{u}'=\myvec{u}+2 \myvec{s}(\myvec{u},t)\,\Delta t$
\citep{ZHANG2006}.  For the gravity source term
\eqref{eq:gravity_term}, this leads to the timestep limit
\begin{align}
\Delta t^K_{\text{grav}}\le\frac{1}{\sqrt{2\gamma(\gamma-1)}}\frac{c}{|\myvec{\nabla}\Phi|}.
\end{align}
The actual timestep has then to be chosen as the minimum of the
hydrodynamical and gravity timesteps.

\subsection{Angular momentum conservation}
\label{sec:ang_mom}

A welcome side effect of the computation of higher order moments of
the solution in DG schemes is that angular momentum is inherently
conserved. Without loss of generality, we show this conservation property
explicitly for the two-dimensional case ($z=0$).  In this case,
the angular momentum density is defined as
\begin{align}
L=x\rho v_y-y\rho v_x,
\end{align}
and the flux momentum tensor in 2D is given by
\begin{align}
\begin{pmatrix}
f_{1,2} & f_{2,2} \\
f_{1,3} & f_{2,3} \\
\end{pmatrix}
=
\begin{pmatrix}
 p+\rho v_1^2 & \rho v_1 v_2 \\
\rho v_1 v_2 & p+\rho v_2 ^2 \\
\end{pmatrix}.
\end{align}
The conservation law for angular momentum can be conveniently derived
from the Euler equations \eqref{eq:euler}. Multiplying the
$x$-momentum equation by $y$ and the $y$-momentum equation by $x$,
applying the product rule and subsequently subtracting the two
equations gives
\begin{align}
\frac{\partial L}{\partial t}+\frac{\partial}{\partial x}(x f_{1,3}-y f_{1,2})+\frac{\partial}{\partial y} (x f_{2,3} - y f_{2,2}) = 0.
\end{align}
In order to obtain the angular momentum conservation law on a cell
basis, this can be integrated over element $K$, resulting in
\begin{align}
  \frac{\mathrm{d}}{\mathrm{d}t}\int_K\!L\,\mathrm{d}V+\int_{\partial K}x(f_{1,3}n_1+f_{2,3}n_2)-y(f_{1,2}n_1+f_{2,2}n_2)\,\mathrm{d}S=0.
\label{eq:angmom_integrated}
\end{align}

On the other hand, for a DG scheme with order $p>1$, we can choose the
test function to be $\phi^K=y$ in the $x$-momentum equation and
$\phi^K=x$ in the $y$-momentum equation in the weak formulation
\eqref{eq:dg_projected} of the Euler equations:
\begin{align}
\frac{\mathrm{d}}{\mathrm{d}t}\int_K\!\rho v_1 y\,\mathrm{d}V-\int_K\!f_{2,2}\,\mathrm{d}V+\int_{\partial K}\numfluxcomp_2 y\,\mathrm{d}S=0,\\
\frac{\mathrm{d}}{\mathrm{d}t}\int_K\!\rho v_2 x\,\mathrm{d}V-\int_K\!f_{1,3}\,\mathrm{d}V+\int_{\partial K}\numfluxcomp_3 x\,\mathrm{d}S=0,
\end{align}
where $\numfluxcomp_2$ and $\numfluxcomp_3$ are the momentum components of the numerical flux function.
By subtracting the above equations and using $f_{1,3}=f_{2,2}=\rho v_1 v_2$ we get the angular momentum equation of
DG, viz.
\begin{align}
\frac{\mathrm{d}}{\mathrm{d}t}\int_K\!L y\,\mathrm{d}V+\int_{\partial K}(\numfluxcomp_3 x - \numfluxcomp_2 y)\,\mathrm{d}S.
\end{align}
This equation is consistent with the exact equation
\eqref{eq:angmom_integrated}, and hence DG schemes of at least second order
accuracy are angular momentum conserving.  However, there is one
caveat to this inherent feature of DG.  In non-smooth regions of the
solution, a limiting scheme has to be applied, which can slightly
modify the angular momentum within a cell and hence lead to a
violation of manifest angular momentum conservation. This is also the
case for the simple limiters we shall adopt and describe in the
subsequent section.

\section{Slope limiting}
\label{sec:slope_limiting}

The choice of the slope-limiting procedure can have a large effect on
the quality of a hydro scheme, as we will demonstrate in
Section~\ref{sec:shock_tube}. Often different limiters and
configurations represent a trade-off between dissipation and
oscillations, and furthermore, the optimal slope limiter is highly
problem dependent. Consequently, the challenge consists of finding a
limiting procedure which delivers good results for a vast range of
test problems. For the DG scheme, this proves to be even more
difficult. The higher order terms of the solution should be discarded
at shocks and contact discontinuities if needed, while at the same
time no clipping of extrema should take place in smooth regions, such that the
full benefit of the higher order terms is ensured. In what follows, we
discuss two different approaches for limiting the linear terms of the
solution as well as a positivity limiter which asserts non-negativity
of density and pressure.

\subsection{Component-wise limiter}
\label{sec:comp_lim}

In order to reduce or completely avoid spurious oscillations, we have
to confine possible over- and undershootings of the high order
solution of a cell at cell boundaries compared to the cell average of
neighbouring cells.  For that purpose, the weights ${w}^K_{2,i}, {w}^K_{3,i}, {w}^K_{4,i}$
which are proportional to the slopes in the $x\text{-}$,
$y\text{-}$, and $z\text{-}$directions, respectively, are limited by comparing them to
the difference of cell-averaged values, viz.
\begin{align}
\nonumber\tilde{w}^K_{2,i}&=\frac{1}{\sqrt{3}}\text{minmod}\left(\sqrt{3}w^K_{2,i}, \beta(w^K_{1,i}-w^{W_K}_{1,i}),\beta(w^{E_K}_{1,i}-w^K_{1,i})\right),\\
\tilde{w}^K_{3,i}&=\frac{1}{\sqrt{3}}\text{minmod}\left(\sqrt{3}w^K_{3,i},\beta(w^K_{1,i}-w^{S_K}_{1,i}),\beta(w^{N_K}_{1,i}-w^K_{1,i})\right),
\label{eq:comp_limiter}\\
\nonumber\tilde{w}^K_{4,i}&=\frac{1}{\sqrt{3}}\text{minmod}\left(\sqrt{3}w^K_{4,i},\beta(w^K_{1,i}-w^{B_K}_{1,i}),\beta(w^{T_K}_{1,i}-w^K_{1,i})\right).
\end{align}
Here, $\tilde{w}^K_{2,i}, \tilde{w}^K_{3,i}, \tilde{w}^K_{4,i}$ are the new weights, $W_K, E_K, S_K, N_K, B_K, T_K$ denote 
cell neighbours in the directions west, east, south, north, bottom, and top, respectively, and the minmod-function 
is defined as
\begin{align}
\text{minmod}(a,b,c)=
\begin{cases}
s\,\text{min}(|a|,|b|,|c|)&s=\text{sign}(a)=\text{sign}(b)=\text{sign}(c)\\
0&\text{otherwise}.
\end{cases}
\end{align}

Each component of the conserved variables, $i=1,\ldots,5$, is limited
separately and hence treated independently.  The $\sqrt{3}$-factors in
equation~\eqref{eq:comp_limiter} account for the scaling of the
Legendre polynomial $\tilde{P}_1(\xi)$, and the parameter
$\beta\in[0.5,1]$ controls the amount of limiting.  The choice of
$\beta=0.5$ corresponds to a total variation diminishing (TVD) scheme
for a scalar problem but introduces more diffusion compared to
$\beta=1$.  The latter value is less restrictive and may yield a more
accurate solution.

If the limited weights are the same as the old weights, i.e.
$\tilde{w}^K_{2,i}=w^K_{2,i}$, $\tilde{w}^K_{3,i}=\tilde{w}^K_{3,i}$,
and $w^K_{4,i}=w^K_{4,i}$, we keep the original component of the
solution $u_i^K\left(\myvec{x},t\right)$ of the cell.  Otherwise, the
component is set to
\begin{align}
u_i^K\left(\myvec{x},t\right)=w_{1,i}^K+\tilde{w}_{2,i}^K\phi_2^K+\tilde{w}_{3,i}^K\phi_3^K+\tilde{w}_{4,i}^K\phi_4^K,
\label{eq:comp_limiter_applied}
\end{align}
where terms with order higher than linear have been discarded.  
Alternative approaches to our simple limiter exist in the form of
hierarchical limiters \citep[e.g.][and references therein]{KUZMIN2014}. 
In these schemes the highest order
terms are limited first before gradually lower order terms 
are modified one after another. In this way also higher order terms 
can be kept in the limiting procedure.
Note
that in equation \eqref{eq:comp_limiter_applied} the cell-averaged values do not
change, and thus the conservation of mass, momentum, and energy is
unaffected. However, the limiter may modify the angular momentum
content of a cell, implying that in our DG scheme it is manifestly
conserved only in places where the slope limiter does not
trigger. Clearly, it is a desirable goal for future work to design
limiting schemes which can preserve angular momentum.

\subsection{Characteristic limiter}
\label{sec:char_lim}

An improvement over the component-wise limiting of the conserved
variables can be achieved by limiting the characteristic variables
instead. They represent advected quantities, and for
the Euler equations we can define them locally by 
linearizing about the cell-averaged value.

The transformation matrices $\mathcal{L}_x^K$,
$\mathcal{L}_y^K$, and $\mathcal{L}_z^K$ are formed
from the left eigenvectors of the flux Jacobian matrix based on the mean values of
the conserved variables of cell $K$,
$\bar{\myvec{u}}^K=\myvec{w}_1^K$. We list all matrices in
Appendix~\ref{sec:eigenvectors}.  The slopes of the characteristic
variables can then be obtained by the matrix-vector multiplications
$\myvec{c}_2^K=\mathcal{L}_x^K\myvec{w}_2^K$,
$\myvec{c}_3^K=\mathcal{L}_y^K\myvec{w}_3^K$,
$\myvec{c}_4^K=\mathcal{L}_z^K\myvec{w}_4^K$, where $\myvec{w}_2^K$,
$\myvec{w}_3^K$ and $\myvec{w}_4^K$ denote the slopes of the conserved
variables in the $x-$, $y-$, and $z-$directions, respectively.  The
transformed slopes are limited as in Section~\ref{sec:comp_lim} with
the minmod-limiting procedure, viz.
\begin{align}
\nonumber\tilde{\myvec{c}}_2^K &=\frac{1}{\sqrt{3}}\text{minmod}\left({\sqrt{3}\myvec{c}_2^K, \beta\mathcal{L}_x^K(\myvec{w}_1^K - \myvec{w}_1^{W_K}), \beta\mathcal{L}_x^K(\myvec{w}_1^{E_K} - \myvec{w}_1^K) }\right) ,\\
\nonumber\tilde{\myvec{c}}_3^K &=\frac{1}{\sqrt{3}}\text{minmod}\left({\sqrt{3}\myvec{c}_3^K, \beta\mathcal{L}_y^K(\myvec{w}_1^K - \myvec{w}_1^{S_K}), \beta\mathcal{L}_y^K(\myvec{w}_1^{N_K} - \myvec{w}_1^K) }\right), \\
\tilde{\myvec{c}}_4^K &=\frac{1}{\sqrt{3}}\text{minmod}\left({\sqrt{3}\myvec{c}_4^K, \beta\mathcal{L}_z^K(\myvec{w}_1^K - \myvec{w}_1^{B_K}), \beta\mathcal{L}_z^K(\myvec{w}_1^{T_K} - \myvec{w}_1^K) }\right).
\label{eq:char_limiter}
\end{align}
If the limited slopes of the characteristic variables are identical to
the unlimited ones (i.e.~$\tilde{\myvec{c}}_2^K=\myvec{c}_2^K$,
$\tilde{\myvec{c}}_3^K=\myvec{c}_3^K$, and
$\tilde{\myvec{c}}_4^K=\myvec{c}_4^K$), the original solution is
kept. Otherwise, the new slopes of the conserved variables are
calculated with the inverse transformation matrices
$\mathcal{R}_x^K=(\mathcal{L}_x^K)^{-1}$,
$\mathcal{R}_y^K=(\mathcal{L}_y^K)^{-1}$, and
$\mathcal{R}_z^K=(\mathcal{L}_z^K)^{-1}$ via
$\tilde{\myvec{w}}_2^K=\mathcal{R}_x^K\tilde{\myvec{c}}_2^K$,
$\tilde{\myvec{w}}_3^K=\mathcal{R}_x^K\tilde{\myvec{c}}_3^K$, and
$\tilde{\myvec{w}}_4^K=\mathcal{R}_x^K\tilde{\myvec{c}}_4^K$.  In this
case, the higher order terms of the solution are set to zero and the
limited solution in the cell becomes
\begin{align}
\myvec{u}^K\left(\myvec{x},t\right)=\myvec{w}_1^K+\tilde{\myvec{w}}_2^K\phi_2^K+\tilde{\myvec{w}}_3^K\phi_3^K+\tilde{\myvec{w}}_4^K\phi_4^K.
\end{align}
The difference between the component-wise limiting of the conserved
variables (Section~\ref{sec:comp_lim}) and the limiting of the more
natural characteristic variables is demonstrated with a shock tube
simulation in Section~\ref{sec:shock_tube}.

\subsection{Total variation bounded limiting}

The limiters discussed so far can effectively reduce overshootings and
oscillations; however, they can potentially also trigger at smooth
extrema and then lead to a loss of higher order
information. Considering the goals of a higher order DG scheme, this
is a severe drawback that can negatively influence the convergence
rate of our DG code. In order to avoid a clipping of the solution at
smooth extrema, the minmod limiter in Sections~\ref{sec:comp_lim} and
\ref{sec:char_lim} can be replaced by a bounded version \citep{CS5}, viz.
\begin{align}
\text{minmodB}(a,b,c)=
\begin{cases}
a&\text{if}\,|a|\le M(\Delta x^K)^2 \\
\text{minmod}(a,b,c)&\text{otherwise}.
\end{cases}
\end{align}
Here, $M$ is a free parameter which is related to the second
derivative of the solution. The ideal choice for it can vary for
different test problems. Furthermore, in the above ansatz, the amount
of limiting depends on the resolution since $|a|\propto\Delta x^K$,
which is not the case for the right hand side of the inequality.  For
reasons of simplicity and generality, we would however like to use
fixed limiter parameters without explicit resolution dependence for
the tests presented in this paper. Hence we define
$\tilde{M}=M \Delta x^K$, and use a constant value for $\tilde{M}$ to
control the strength of the bounding applied to the minmod
limiter. With the minmod-bounded limiting approach, the high accuracy
of the solution in smooth regions is retained while oscillations,
especially in post-shock regions, can be eliminated.

\subsection{Positivity limiting}
\label{sec:positivity_limiter}

When solving the equations of hydrodynamics for extreme flows, care
has to be taken in order to avoid negative pressure or density values
within the cells and at cell interfaces.  A classical example is high
Mach number flows, for which in the pre-shock region the total energy
is dominated by the kinetic energy. Because the pressure is calculated
via the difference between total and kinetic energies, it can easily become
negative in numerical treatments without special precautions.  

The use of an ordinary slope limiter tends to be only of limited help
in this situation, and only delays the blow-up of the solution. In
fact, it turns out that even with TVD limiting and arbitrarily small
timesteps, it is not guaranteed in general that unphysical negative
values are avoided.  Nevertheless, it is possible to construct
positivity preserving FV and DG
schemes, which are accurate at the same time. Remarkably, the
latter means that high order accuracy can be retained in smooth
solutions and furthermore the total mass, momentum, and energy in each
cell are conserved.  For our DG code, we adopt the positivity limiting
implementation following \citet{ZHANG2010}.

For constructing this limiter, a quadrature rule including the
boundary points of the integration interval is needed. One possible
choice consists of $m$-point Gauss-Lobatto-Legendre (GLL) quadrature
rules (Appendix \ref{sec:gll_quadrature}), which are exact for polynomials of degree $k\le2m-3$.
Let $\xi^{\oned}_{1}<\ldots<\xi^{\oned}_{k+1}\in[-1,1]$ be the
one-dimensional Gauss quadrature points and
$\hat{\xi}^{\oned}_{1}<\ldots<\hat{\xi}^{\oned}_{m}\in[-1,1]$ the GLL
quadrature points.  Quadrature rules for the domain $[-1,-1]^3$ which
include the interface Gauss quadrature points can be constructed by
tensor products of Gauss and GLL quadrature points:
\begin{align}
\nonumber &S_x=\{(\hat{\xi}^{\oned}_r, \xi^{\oned}_s, \xi^{\oned}_t): 1\le r \le m, 1\le s \le k+1, 1\le t \le k+1\},\\
\nonumber &S_y=\{(\xi^{\oned}_r, \hat{\xi}^{\oned}_s, \xi^{\oned}_t): 1\le r \le k+1, 1\le s \le m, 1\le t \le k+1\},\\
&S_z=\{(\xi^{\oned}_r, \xi^{\oned}_s, \hat{\xi}^{\oned}_t): 1\le r \le k+1, 1\le s \le k+1, 1\le t \le m\}.
\label{eq:union_quad_points}
\end{align}
The union $S=S_x \cup S_y \cup S_z$ of these sets constitutes
quadrature points of a rule which is exact for polynomials of degree
$k$ and furthermore contains all the points for which we calculate
fluxes in equation~\eqref{eq:dg_term3}.  It can be shown that if the
solution is positive at these union quadrature points, the solution
averaged after one explicit Euler step will stay positive if a
positivity preserving flux calculation and an adequate timestep are
used.  

The positivity limiter is operated as follows.  For a DG scheme with
polynomials of maximum order $k$, choose the smallest integer $m$ such
that $m\ge (k+3)/2$.  Carry out the following computations for every
cell $K$. First, determine the minimum density at the union quadrature
points,
\begin{align}
\rho_{\text{min}}^K=\min_{\myvec{\xi}\in S} \rho^K(\myvec{\xi}).
\end{align}
Use this minimum density for calculating the factor
\begin{align}
\theta_1^K=\min\left\{\left|\frac{\bar{\rho}^K-\epsilon}{\bar{\rho}^K-\rho_{\text{min}}^K}\right|,1\right\}, 
\end{align}
where $\epsilon\approx 10^{-10}$ is a small number representing the
target floor for the positivity limiter.  Then, modify the higher order 
terms of the density by multiplying the corresponding weights
with the calculated factor,
\begin{align}
w_{j,1}^K\leftarrow\theta_1^K w_{j,1}^K,\quad j=2,\ldots,N(k).
\end{align}
At this point, the density at the union quadrature points is positive
($\ge \epsilon$), and as desired, the mean density
$\bar{\rho}^K=w_{1,1}^K$ and therefore the total mass has not been
changed.

In order to enforce pressure positivity at the union quadrature points
$\myvec{\xi}\in S$, we compute the factor
\begin{align}
\theta_2^K=\min_{\myvec{\xi}\in S} \tau^K(\myvec{\xi}),
\end{align}
with
\begin{align}
\tau^K(\myvec{\xi})=
\begin{cases}
1\quad&\text{if}\,\,\,p^K(\myvec{\xi})\ge \epsilon \\
\tau_\ast\quad&\text{such that}\,\,\,p^K(\myvec{u}^K(\myvec{\xi})+\tau_\ast(\myvec{u}^K(\myvec{\xi})-\bar{\myvec{u}}^K))=\epsilon,
\end{cases}
\label{eq:p_positivity}
\end{align}
and limit all higher order terms ($j>=2$) for all components of the state vector:
\begin{align}
w_{j,i}^K\leftarrow\theta_2^K w_{j,i}^K,\quad j=2,\ldots,N(k),\quad i=1,\ldots,5.
\end{align}
The idea of the second case of equation~\eqref{eq:p_positivity} is to
calculate a modification factor $\tau_\ast$ for the higher order terms
$\myvec{u}^K(\myvec{\xi})-\bar{\myvec{u}}^K$, such that the new
pressure at the quadrature point equals $\epsilon$. Furthermore, this
second case represents a quadratic equation in $\tau_\ast$, which has
to be solved carefully by minimizing round-off errors. In our
implementation, we improve the solution for $\tau_\ast$ iteratively
with a small number of Newton-Raphson iterations.  

The outlined method ensures positivity of pressure and density of the
mean cell state after the subsequent timestep under several
conditions. First, a positivity preserving flux calculation has to
be used, e.g.~the Godunov flux or the Harten-Lax-van Leer flux 
is suitable.  Secondly, when adopting an RK method, it should be strong
stability preserving (SSP); these methods are convex combinations of
explicit Euler methods, and hence a number of relevant properties and
proofs valid for Euler's method also hold for these higher order time
integrators.  Lastly, the local timestep for the DG scheme has to be
set to
\begin{multline}
\Delta t^K=\text{CFL}\cdot\min\left\{\frac{1}{2k+1}, \frac{\hat{\omega}_1^\oned}{2}\right\}\\
\left(\frac{|v_1^K|+c^K}{\Delta x_1^K}+\frac{|v_2^K|+c^K}{\Delta x_2^K}+\frac{|v_3^K|+c^K}{\Delta x_3^K}\right)^{-1},
\label{eq:time_step_pos}
\end{multline}
where $\hat{\omega}_1^\oned$ is the first GLL weight. Compared to \citet{ZHANG2010}
we obtain an additional factor of $1/2$ since our reference domain is
$[-1,1]$ and therefore the sum of the GLL weights is
$\sum_{q=1}^{m}\hat{\omega}_q^\oned=2$.  The first GLL weight is
$\hat{\omega}_1^\oned=1$ for our second order DG (DG-2) method ($k=1$) and
$\hat{\omega}_1^\oned=1/3$ for our third order and fourth order method ($k=2,3$).
Depending on the order, the timestep with positivity limiting can be
slightly more restrictive. We conclude this section by remarking that
the positivity limiting procedure is a local operation and does not
introduce additional inter-cell communication in the DG scheme.

\begin{figure}
\centering
\includegraphics{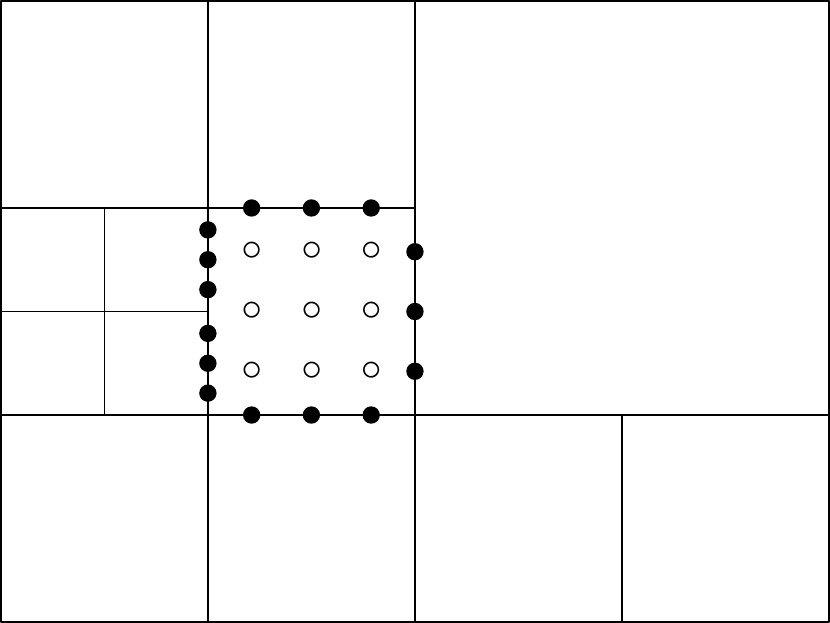}
\caption{Quadrature points of an AMR boundary cell in our 2D, third order 
  ($k=2$) version of the code.  Interfaces with neighbouring
  cells on a finer AMR level are integrated with the quadrature points
  of the smaller cells. In this way no accuracy is lost at AMR
  boundaries and the order of our DG code is unaffected.}
\label{fig:dg_amr}
\end{figure}

\section{DG with AMR}
\label{sec:dg_amr}

Many astrophysical applications involve a high dynamic range in
spatial scales. In grid-based codes, this multi-scale challenge can be
tackled with the AMR technique. In this
approach, individual cells are split into a set of smaller subcells
if appropriate (see Fig.~\ref{fig:ref_coarse} for a sketch of the
two-dimensional case), thereby increasing the resolution locally.  We
adopt a tree-based AMR implementation where cubical cells are split
into eight subcells when a refinement takes place. This allows a
particularly flexible adaption to the required spatial resolution.

Our implementation largely follows that in the {\small RAMSES} code
\citep{RAMSES}. The tree is organized into cells and nodes. The root
node encompasses the whole simulation domain and is designated as
level $l=0$. A node at level $l$ always contains eight children, which
can be either another node on a smaller level $l' = l+1$, or a leave
cell. In this way, the mesh of leave cells is guaranteed to be volume
filling. An example of such an AMR mesh in 2D is shown in
Fig.~\ref{fig:dg_amr}.

To distribute the work load onto many MPI processes, the tree is split
into an upper part, referred to as `top tree' in the following, and
branches that hang below individual top tree nodes. Every MPI process
stores a full copy of the top tree, whereas each of the lower branches
is in principle stored only on one MPI rank. However, some of the
branch data are replicated in the form of a ghost layer around each
local tree to facilitate the exchange of boundary information. The
mesh can be refined and structured arbitrarily, with only one
restriction. The level jump between a cell and its $26$ neighbours has
to be at most $\pm 1$.  This implies that a cell can either have one
or four split cells as direct neighbours in a given direction.  In
order to fulfil this level jump condition, additional cells may have
to be refined beyond those where the physical criterion demands a
refinement. 

To simplify bookkeeping, we store for each cell and node the indices
of the father node and the six neighbouring cells or nodes. In case of
a split neighbour, the index of the node at the same level is stored
instead.  To make the mesh smoother and guarantee a buffer region
around cells which get refined due to the physical refinement
criterion, additional refinement layers are added as needed. In the
AMR simulations presented in this work, we use one extra layer of
refined cells around each cell flagged by the physical criterion for
refinement.

\subsection{Refinement criterion}
\label{sec:ref_criterion}

Many different refinement strategies can be applied in an AMR code,
for example, the refinement and derefinement criterion may aim to keep
the mass per cell approximately constant. Another common strategy for
refinement is to focus on interesting regions such as shocks, contact
discontinuities, or fluid instabilities. In this work, we apply the
mesh refinement simply at locations where the density gradient is
steep. To be precise, cell $K$ is refined if the following criterion
is met:
\begin{align}
\max(w^K_{2,0},w^K_{3,0},w^K_{4,0})>\alpha \cdot w_\mathrm{t}.
\label{eq:cell_ref}
\end{align}
Here, $w^K_{2,0}$, $w^K_{3,0}$, and $w^K_{4,0}$ are the density changes divided by $\sqrt{3}$
along the $x-$, $y-$, and $z-$ directions, respectively.
The refinement is controlled by the target slope parameter $w_\mathrm{t}$ and a range factor $\alpha$. 
We have introduced the latter with the purpose of avoiding an oscillating
refinement-derefinement behaviour.  In this work, we adopt the values
$w_\mathrm{t}=0.01$ and $\alpha=1.1$.  The reason for using the
density changes instead of the physical slopes is that in this way a runaway
refinement can be avoided.  

Leave nodes of the AMR tree are kept refined if the following
equation is true:
\begin{align}
\max(w^L_{2,0},w^L_{3,0},w^L_{4,0})>\frac{1}{\alpha} \cdot w_\mathrm{t}.
\label{eq:node_ref}
\end{align}
The weights on the left hand side are the leave node weights 
calculated with a projection of the eight subcell solutions.
If inequality \eqref{eq:node_ref} does not hold, the node
gets derefined and the eight subcells are merged into one.

\subsection{Mesh refinement}
\label{sec:refinement}
\begin{figure}
\centering
\includegraphics{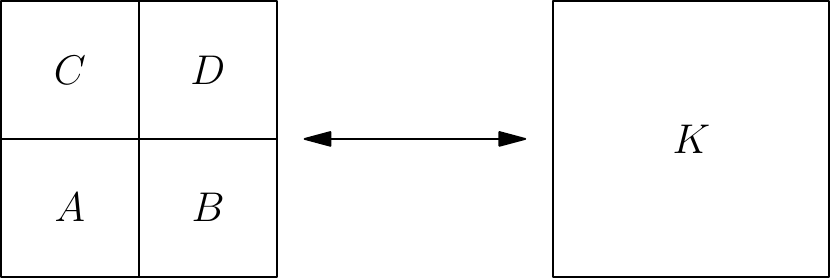}
\caption{Refinement (arrow to the left) and coarsening (arrow to the
  right) of a cell in the 2D version of {\small TENET}. In both
  operations the solution on the new cell structure is inferred by an
  $L^2$-projection of the current polynomial solution. In doing so, no
  information is lost in a refinement, and as much information as
  possible is retained in a derefinement.}
\label{fig:ref_coarse}
\end{figure}
Let $K=\{\myvec{\xi}|\myvec{\xi}\in[-1,1]^3\}$ be the cell which we
want to refine into eight subcells, and $A$, $B$, $C$, $\ldots$ , $H$
denote the daughter cells, as illustrated in Fig.~\ref{fig:ref_coarse}
for the two-dimensional case. The refinement can be achieved without
higher order information loss if the solution polynomial of the coarse
cell is correctly projected onto the subcells. In the
following, we outline the procedure for obtaining the solution on
subcell
$A=\{\myvec{\xi}|\myvec{\xi}\in[-1,0]\times[-1,0]\times[-1,0]\}$, the
calculations for the other subcells are done in an analogous way.  

The weights of the solution on subcell $A$ are obtained by solving the
minimization problem
\begin{align}
\min_{\left\{w^A_{l,i}\right\}_l}\int_{A}\!\left(u_i^K-u_i^A\right)^2\,\mathrm{d}V,\quad i=1,\ldots,5,
\label{eq:ref_min}
\end{align}
where $\myvec{u}^K=\sum\limits_{l=1}^{N(k)}\myvec{w}_l^K\phi_l^K$
is the given solution on the coarse cell and 
$\myvec{u}^A=\sum\limits_{l=1}^{N(k)}\myvec{w}_l^A\phi_l^A$
are the polynomials of the conserved variables on cell $A$ we are looking for.
The minimization ansatz \eqref{eq:ref_min} is solved by projecting
the solution of the coarse cell onto the basis functions of subcell $A$,
\begin{align}
\myvec{w}^A_j=\frac{1}{|A|}\int_A\!\myvec{u}^K\phi^A_j\,\mathrm{d}V,\quad j=1,\ldots,N(k).
\end{align}
We can plug in the solution $\myvec{u}^K$ and move 
the time- but not space-dependent weights in front of the integral,
\begin{align}
w_{j,i}^A=\sum\limits_{l=1}^{N(k)}w_{l,i}^K\frac{1}{|A|}\int_A\!\phi_l^K\phi_j^A\,\mathrm{d}V,\quad i=1,\ldots,5\,,j=1,\ldots,N(k).
\end{align}
If we define the projection matrix 
\begin{align}
(\mathsf{P}_A)_{j,l}=\frac{1}{|A|}\int_A\!\phi_l^K\phi_j^A\,\mathrm{d}V,
\label{eq:ref_pa_integral}
\end{align}
and introduce for each conserved variable $i=1,\ldots,5$ the weight vector
$\hat{\myvec{w}}_i=(w_{1,i}, w_{2,i},\ldots, w_{N,i} )^\top$,
the weights of the subcell solution $\myvec{u}^A$
are simply given by the matrix-vector multiplications
\begin{align}
\hat{\myvec{w}}_i^A=\mathsf{P}_A\hat{\myvec{w}}_i^K\,\quad i=1,\ldots,5.
\label{eq:ref_weights}
\end{align}
The matrix \eqref{eq:ref_pa_integral} can be computed exactly by transforming the integral
to the reference domain of cell $A$,
\begin{align}
(\mathsf{P}_A)_{j,l}=\frac{1}{8}\iiint\displaylimits_{[-1,1]^3}\!\phi_j\left( \frac{\xi_1-1}{2},  \frac{\xi_2-1}{2}, \frac{\xi_3-1}{2}\right)\phi_l(\xi_1,\xi_2,\xi_3)\,\mathrm{d}\myvec{\xi},
\end{align}
and applying a Gaussian quadrature rule with $(k+1)^3$ points.  The
projection integrals for the other subcells ($B$, $C$, $\ldots$, $H$)
are given in Appendix~\ref{app:refinement_matrices}.  The refinement
matrices are the same for all the cells; we calculate them once in the
initialization of our DG code.

\subsection{Mesh derefinement}
\label{sec:derefinement}

\begin{figure*}
\begin{center}
\includegraphics[width=0.45\textwidth]{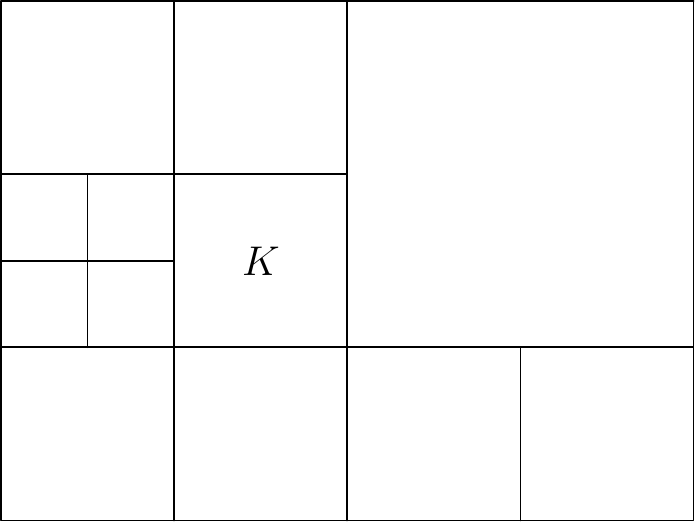} \hspace{5mm}
\includegraphics[width=0.45\textwidth]{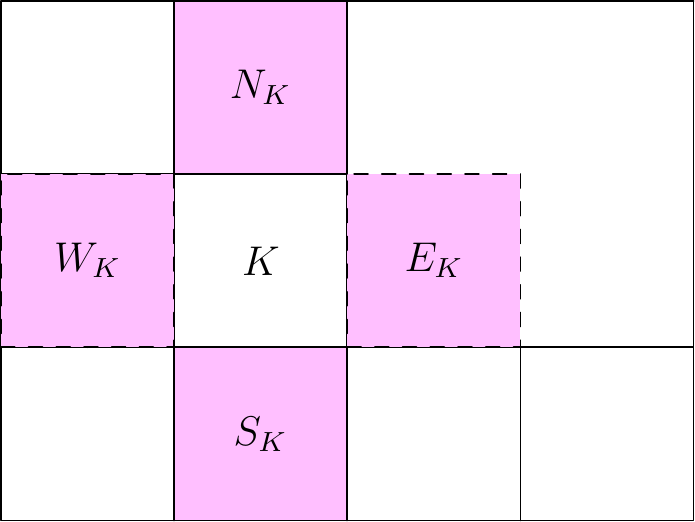}
\caption{Definition of neighbours in the slope-limiting procedure at
  AMR boundaries, shown for the 2D version of {\small TENET}.  The
  neighbours of cell $K$ on the left are on a finer level. For the
  slope limiting of cell $K$, the node weights are used, which are
  calculated by projecting the solutions of the subcells onto the
  encompassing node volume $W_K$. The right neighbour of cell $K$ on
  the other hand is here a coarser cell; in this case, the solution of
  this neighbour has to be projected onto the smaller volume $E_K$. In
  the slope limiting routine of {\small TENET}, this is only done in a
  temporary fashion without actually modifying the mesh.  }
\label{fig:adaplim}
\end{center}
\end{figure*}

When the eight cells $A$, $B$, $\ldots$, $H$ are merged into one coarser
cell $K$, some information is unavoidably lost.  Nevertheless, in
order to retain as much accuracy as possible, the derefinement should
again be carried out by means of an $L^2$-projection,
\begin{align}
\nonumber\min_{\left\{w^K_{l,i}\right\}_l}\int_{A}\!\left(u_i^K-u_i^A\right)^2\,\mathrm{d}V+\int_{B}\!\left(u_i^K-u_i^B\right)^2\,\mathrm{d}V+
  \ldots +\\
\int_{H}\!\left(u_i^K-u_i^H\right)^2\,\mathrm{d}V,\quad i=1, \ldots, 5.
\end{align}
Here, $\myvec{u}^A$, $\myvec{u}^B$, $\ldots$, $\myvec{u}^H$ denote the given solutions on the cells to be derefined,
and $\myvec{u}^K$ is the solution on the coarser cell we are looking for.
The minimization problem is solved by the weights
\begin{align}
\nonumber\myvec{w}_j^K=\frac{1}{|K|}\left(\int_{A}\!\myvec{u}^A\phi_j^K\,\mathrm{d}V+\int_{B}\!\myvec{u}^B\phi_j^K\,\mathrm{d}V+
  \ldots +\int_{H}\!\myvec{u}^H\phi_j^K\,\mathrm{d}V\right),\\
\quad j=1, \ldots , N(k).
\end{align}
We insert the solutions on the cells $A$, $B$, $\ldots$, $H$ and use the definition of the refinement matrices,
\begin{align}
\nonumber w_{j,i}^K&=\frac{1}{|K|}\left(\int_A\!\sum\limits_{l=1}^{N(k)}w_{l,i}^A\phi_l^A\phi_j^K\,\mathrm{d}V+\ldots+\int_H\!\sum\limits_{l=1}^{N(k)}w_{l,i}^H\phi_l^H\phi_j^K\,\mathrm{d}V\right)\\
\nonumber&=\frac{1}{|K|}\left(\sum\limits_{l=1}^{N(k)}w_{l,i}^A|A|(\mathsf{P}_A^\top)_{j,l}+\ldots+\sum\limits_{l=1}^{N(k)}w_{l,i}^H|H|(\mathsf{P}_H^\top)_{j,l}\right),\\
\quad &i=1,\ldots,5\;\;\;,\;\;\;j=1,\ldots,N(k).
\end{align}
If we again use the vector $\hat{\myvec{w}}_i^K$ consisting of the weights of the solution for the conserved variable $i$, 
the weights of the solution on cell $K$ can be computed with the transposed refinement matrices,
\begin{align}
\hat{\myvec{w}}_i^K=\frac{1}{8}\left(\mathsf{P}_A^\top\hat{\myvec{w}}_i^A+\mathsf{P}_B^\top\hat{\myvec{w}}_i^B+\ldots+\mathsf{P}_H^\top\hat{\myvec{w}}_i^H\right),\quad i=1,\ldots,5.
\label{eq:deref_weights}
\end{align}
Here we have used the fact that the cells to be merged have the same volume, 
$|A|=|B|=\ldots=|H|=\frac{1}{8}|K|$.

The refinement and derefinement matrices are orthogonal, 
i.e. $\mathsf{P}_A\mathsf{P}_A^\top=\mathsf{I}$.
Thus, if a cell is refined and immediately thereafter derefined, 
the original solution on the coarse cell is restored.
This property is desirable for achieving a stable AMR algorith;
in our approach it can be shown explicitly by inserting the subcell weights 
from equation \eqref{eq:ref_weights} of every subcell in equation \eqref{eq:deref_weights}.
While the refinement of eight cells into a subcell preserves the 
exact shape of the solution, this is in general not the
case for the derefinement. 
For example, if the solution is 
discontinuous across the eight subcells, it cannot
be represented in a polynomial basis on the coarse cell.
After derefining we limit the solution before further 
calculations are carried out, especially for asserting positivity.

\subsection{Limiting with AMR}
\label{seq:amr_limiting}

In the case of AMR boundary cells, the limiting procedure has to be
well defined. For the limiting of cell $K$, the slope limiters
described in Sections~\ref{sec:comp_lim} and \ref{sec:char_lim} need
the average values of the neighbouring cells in each direction in
order to adjust the slope of the conserved variables.  However, if the
cell neighbours are on a different AMR level, they are smaller or
larger compared to the cell to limit, and a single neighbouring cell
in a specific direction is not well defined.  

Fortunately, there is a straightforward way in DG to cope with this
problem.  If a neighbouring cell is on a different level, the
polynomials of the neighbour are projected onto a volume which is
equal to the volume of the cell to refine, see
Fig.~\ref{fig:adaplim}. This can be done with the usual refinement and
derefinement operations as described in Sections~\ref{sec:refinement}
and \ref{sec:derefinement}. By doing so, the limiting at AMR
boundaries can be reduced to the limiting procedure on a regular grid.
In AMR runs, we also slightly adjust the positivity limiting scheme.
For cells which have neighbours on a higher level, the positivity
limiter is not only applied at the locations of the union quadrature
points (equation \ref{eq:union_quad_points}), but additionally at the
quadrature points of the cell interfaces to smaller neighbours.  By
doing so, negative values in the initial conditions of the Riemann
problems are avoided, which have to be solved in the integration of
these interfaces.

\subsection{Main simulation loop}
\label{sec:main_loop}

Our new {\small TENET} code has been developed as an extension of the
{\small AREPO} code \citep{AREPO}, allowing us to reuse {\small
  AREPO}'s input-output routines, domain decomposition, basic tree
infrastructure, neighbour search routines, and gravity solver. This
also helps to make our DG scheme quickly applicable in many science
applications. We briefly discuss the high-level organization of our
code and the structure of the main loop in a schematic way, focusing
on the DG part.

\begin{enumerate}
\setlength\itemsep{1em}
\item Compute and store quadrature points and weights,\\
\phantom{\quad,\quad} basis function values, and projection matrices.
\item Set the initial conditions by means of an $L^2$-projection.
\item Apply slope limiter.
\item Apply positivity limiter. 
\item While $t < t_\text{max}$:
\begin{enumerate}
\setlength\itemsep{0.5em}
\item Compute timestep.
\item For every RK stage:
\begin{enumerate}
\setlength\itemsep{0.3em}
\item Calculate $\myvec{R}_K$ of the differential equation \eqref{eq:spatial_discretisation}\\
\phantom{\quad\quad\,\,\,\,}(inner and outer integral).
\item Update solution to the next RK stage.
\item Update node data.
\item Apply slope limiter.
\item Apply positivity limiter. 
\end{enumerate}
\item Do refinement and derefinement.
\item $t = t + \Delta t$.
\end{enumerate}
\end{enumerate}

The basis function values and the quadrature data are computed in a
general way for arbitrary spatial order as outlined in
Appendices~\ref{sec:legendre_polynomials},
\ref{sec:gaussian_quadrature}, and \ref{sec:gll_quadrature}.  For the
time integration, our code can be provided with a general Butcher
tableau (Table \ref{tab:butcher}), specifying the desired RK method.
By keeping these implementations general, the spatial order and time
integrator can be changed conveniently. 

The first step of an RK stage consists of calculating $\myvec{R}_K$ of
the differential equation \eqref{eq:spatial_discretisation}, including
possible source terms.  This involves computing the inner integral by
looping over the cells and the outer integral by looping over the cell
interfaces.  After updating the solution weights for every cell, the
hydrodynamical quantities on the AMR nodes have to be updated, such
that they can be subsequently accessed during the slope-limiting
procedure (Section~\ref{seq:amr_limiting}).  After the RK step, all
cells are checked for refinement and derefinement, and the mesh is
adjusted accordingly.

\begin{figure*}
\centering
\includegraphics{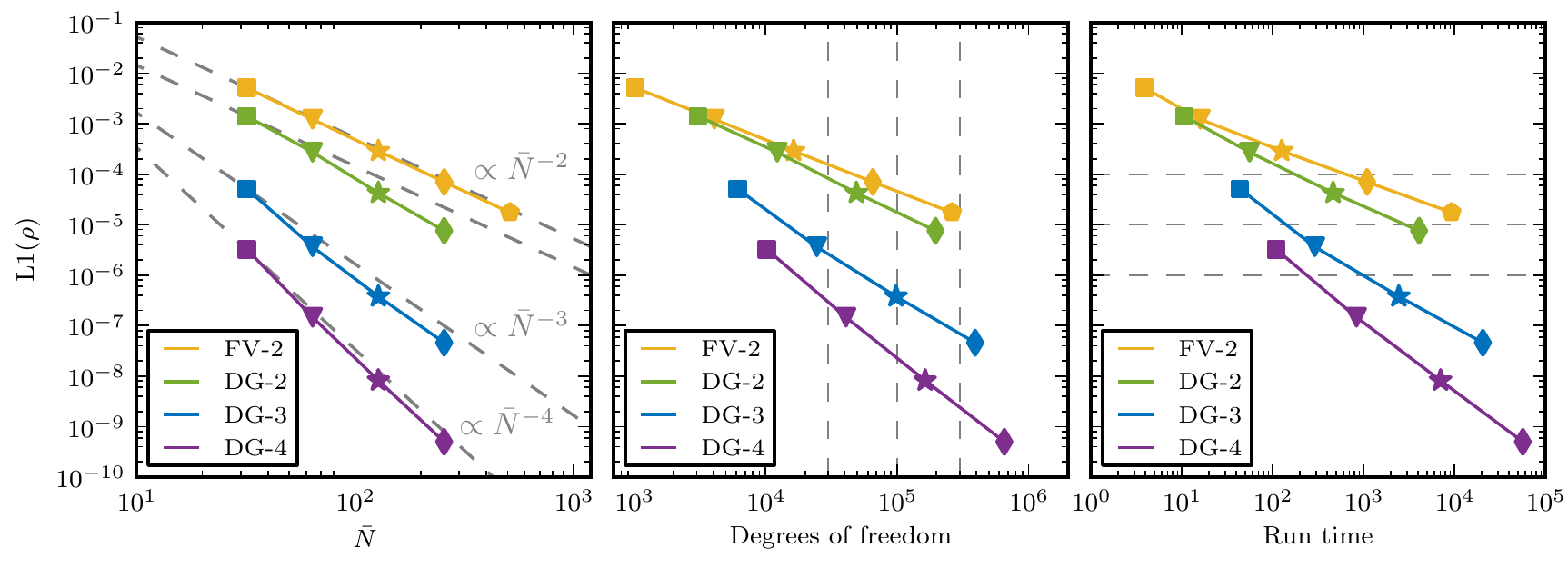}
\caption{Left-hand panel: L1 error norm as a function of linear
  resolution for the two-dimensional isentropic vortex test.  Each
  data point corresponds to a simulation, and different colours indicate
  the different methods.  We find a convergence rate as expected
  (dashed line) or slightly better for all schemes, indicating the
  correctness of our DG implementation.
  Middle panel: the same simulation errors as a function of degrees of freedom (DOF),
  which is an indicator for the memory requirements.
  For our test runs the higher order DG methods are more accurate at the same number of DOF.
  Right-hand panel: L1 error
  norm versus the measured run time of the simulations. The second order 
  FV implementation (FV-2) and the second order DG (DG-2) realization
  are approximately equally efficient in this test, i.e.~a given
  precision can be obtained with a similar computational cost.  In
  comparison, the higher order methods can easily be faster by more
  than an order of magnitude for this smooth problem. This illustrates
  the fact that an increase of order ($p$-refinement) of the numerical
  scheme can be remarkably more efficient than a simple increase of
  grid resolution ($h$-refinement).  }
\label{fig:yee_vortex}
\end{figure*}

\section{Validation}
\label{sec:validation}

In this section, we discuss various test problems which are either
standard tests or chosen for highlighting a specific feature of the DG
method.  For most of the test simulations, we compare the results to a
traditional second order FV method (FV-2). For definiteness, we use the
{\small AREPO} code to this end, with a fixed Cartesian grid and its
standard solver as described in \citet{AREPO}. The latter consists of
a second order unsplit Godunov scheme with an exact Riemann solver and
a non-TVD slope limiter.  Recently, some modifications to the FV
solver of {\small AREPO} have been introduced for improving its
convergence properties \citep{AREPO_CONVERGENCE} when the mesh is
dynamic. However, for a fixed Cartesian grid, this does not make a
difference and the old solver used here performs equally well.  

There are several important differences between FV and DG methods. In
an FV scheme, the solution is represented by piecewise constant states,
whereas in DG the solution within every cell is a polynomial
approximation. Moreover, in FV a reconstruction step has to be carried
out in order to recreate higher order information. Once the states at
the interfaces are calculated, numerical fluxes are computed and the
mean cell values updated.  In the DG method, no higher order
information is discarded after completion of a step and therefore no
subsequent reconstruction is needed. DG directly solves also for the
higher order moments of the solution and updates the weights of the
basis functions in every cell accordingly.  

For all the DG tests presented in this section, we use an SSP RK time
integrator of an order consistent with the spatial discretization, and
the fluxes are calculated with the HLLC approximate Riemann solver.
Furthermore, if not specified otherwise, we use for all tests the
positivity limiter in combination with the characteristic limiter in
the bounded version with the parameters $\beta=1$ and
$\tilde{M}=0.5$. Specifically, we only deviate from this configuration
when we compare the limiting of the characteristic variables to the
limiting of the conserved variables in the shock tube test in
Section~\ref{sec:shock_tube}, and when the angular momentum
conservation of our code is demonstrated in
Section~\ref{sec:keplerian_disc} with the cold Keplerian disc problem.
The higher order efficiency of our DG code is quantified in the test
problem of Section~\ref{sec:yee_vortex}, and the 3D version of our
code is tested in a Sedov-Taylor blast wave simulation in
Section~\ref{sec:sedov_taylor}. In Section~\ref{sec:square_advection},
we show that advection errors are much smaller for DG compared to FV-2.
Finally, the AMR capabilities of {\small TENET} are illustrated with a
high-resolution Kelvin-Helmholtz (KH) instability test in
Section~\ref{sec:kh}.

\subsection{Isentropic vortex}
\label{sec:yee_vortex}

\begin{figure*}
\centering
\includegraphics{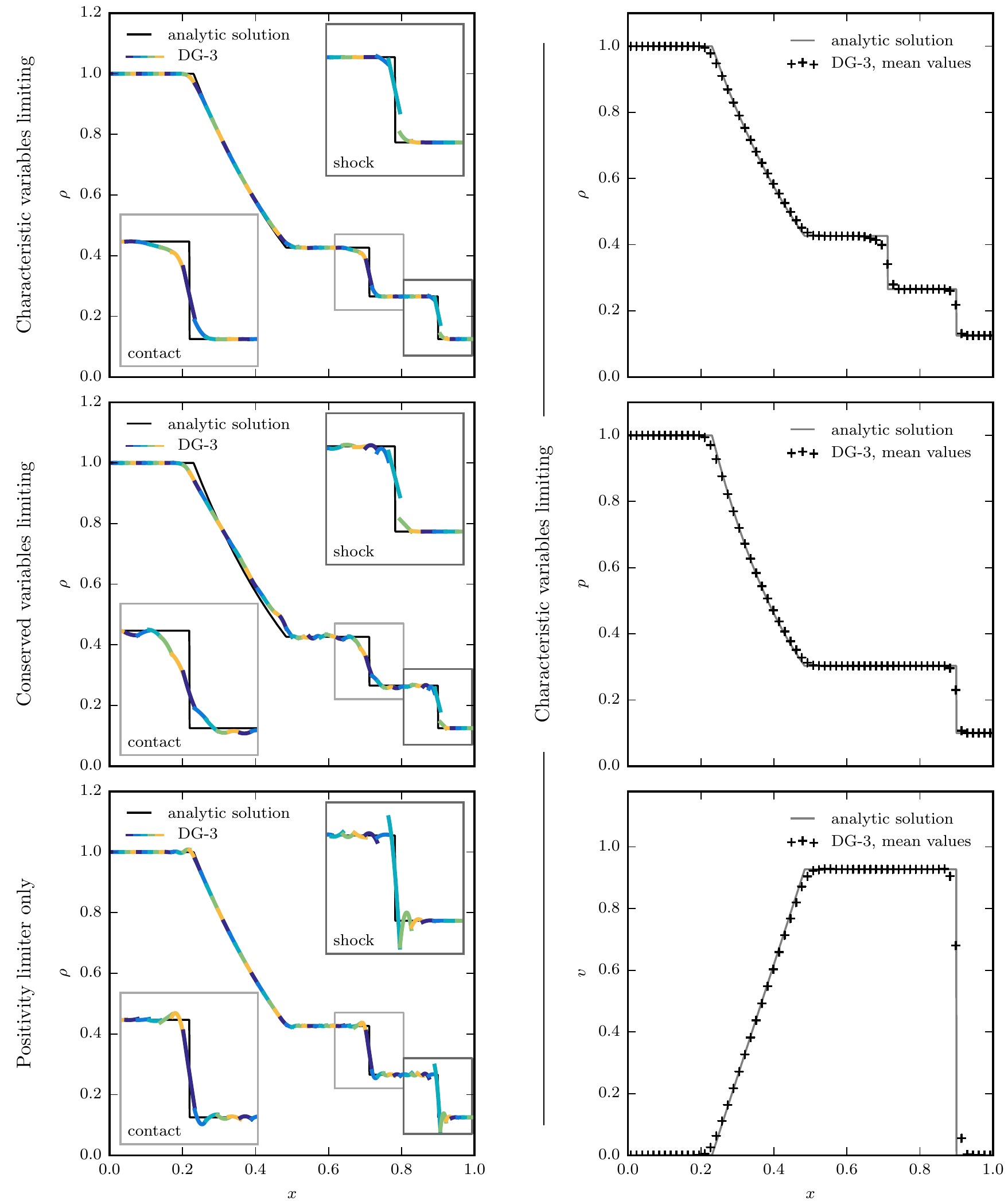}
\caption{Left-hand panels: Sod shock tube problem calculated with
  DG-3 and different limiting approaches.  We show the full
  polynomial DG solutions of the density, where the different colours
  correspond to different cells.  The best result is achieved when the
  limiting is carried out based on the characteristic variables. As
  desired, the numerical solution is discontinuous and of low order at
  the shock. If the conserved variables are limited instead, the
  obtained solution is much less accurate, including numerical
  oscillations. The bottom-left panel shows the result without a slope
  limiter. The solution is of third order in every cell (parabolas)
  leading to an under- and overshooting at the shock as well as
  spurious oscillations. Right-hand panels: comparison of the mean
  cell values with the analytic solution for the characteristic
  variables limiter.  Advection errors wash out the solution at the
  contact rapidly, until it can be represented smoothly by
  polynomials.  Overall, we find very good general agreement with
  the analytic solution, especially at the position of the shock.  }
\label{fig:shock_tube}
\end{figure*}

In this first hydrodynamical test problem, we verify the correctness
of our DG implementation by measuring the convergence rate towards the
analytic solution for different orders of the scheme.  Additionally,
we investigate the precision of the FV-2 and DG schemes as a function of
computational cost.  

An elementary test for measuring the convergence of a hydrodynamical
scheme is the simulation of one-dimensional travelling waves at
different resolutions \citep[e.g.][]{ATHENA}. However, the DG scheme
performs so well in this test, especially at higher order, that the
accuracy is very quickly limited by machine precision, making the
travelling sound wave test impractical for convergence studies of our
DG implementation. We hence use a more demanding setup, the stationary
and isentropic vortex in two dimensions \citep{YEE_1999}. The
primitive variables density, pressure, and velocities in the initial
conditions are
\begin{align*}
p(r)&=\rho^\gamma,\\
\rho(r)&=\left[1-\frac{(\gamma-1)\beta^2}{8\gamma\pi^2}\exp(1-r^2)\right]^\frac{1}{\gamma-1},\\
v_x(r)&=-(y-y_0)\frac{\beta}{2\pi}\exp\left(\frac{1-r^2}{2}\right),\\
v_y(r)&=(x-x_0)\frac{\beta}{2\pi}\exp\left(\frac{1-r^2}{2}\right),
\end{align*}
with $\beta=5$, and the adiabatic index $\gamma=7/5$.  With this
choice for the primitive variables, the centrifugal force at each
point is exactly balanced by the pressure gradient pointing towards
the centre of the vortex, yielding a time-invariant situation.

The vortex is smooth and stationary, and every change during the time
integration with our numerical schemes can be attributed mainly on
numerical truncation errors.  In order to break the spherical symmetry
of the initial conditions, we additionally add a mild velocity boost
of $v_\text{b}=(1,1)$ everywhere.  The simulation is carried out in
the periodic domain $(x,y)\in\left[0,10\right]^2$ with the centre of
the vortex at $(x_0,y_0)=(5,5)$ and run until $t=10$, corresponding to
one box crossing of the vortex.  We compare the obtained numerical
solution with the analytic solution, which is given by the initial
conditions.  The error is measured by means of the L1 density norm,
which we define for the FV method as
\begin{align}
\mathrm{L}1=\frac{1}{N}\sum\limits_{i}|\rho'_i-\rho_i^0|,
\label{eq:L1_norm_fv}
\end{align}
where $N$ is the total number of cells, and $\rho'_i$ and $\rho_i^0$
denote the density in cell $i$ in the final state and in the initial
conditions, respectively.  This norm should be preferred over the L2
norm, since it is more restrictive.  

For DG the error is inferred by calculating the integral
\begin{align}
\mathrm{L}1=\frac{1}{V}\int_V\!\ |\rho'(x,y)-\rho^0(x,y)|\,\mathrm{d}V,
\label{eq:L1_norm}
\end{align}
with the density solution polynomial $\rho'(x,y)$ and the analytic
expression $\rho^0(x,y)$ of the initial conditions.  We compute the
integral numerically with accuracy of order $p+2$, where $p$ is the
order of the DG scheme. In this way, we assert that the error
measurement cannot be dominated by errors in the numerical
integration of the error norm.

The result of our convergence study is shown in the left-hand panel of
Figure~\ref{fig:yee_vortex}. For every method we run the simulation
with several resolutions, indicated by different symbols. The L1
errors decrease with resolution and meet the expected convergence
rates given by the dashed lines, pointing towards the correct
implementation of our DG scheme. At a given resolution, the
DG-2 code achieves a higher precision compared to the FV-2 code, 
reflecting the larger number of flux calculations involved in
DG. The convergence rates are equal, however, as expected. The higher order 
DG formulations show much smaller errors, which also drop more
rapidly with increasing resolution.

In the middle panel of Fig.~\ref{fig:yee_vortex}, we show the same plot
but substitute the linear resolution on the horizontal axis with the
number of degrees of freedom (DOF). This number is closely related to 
the used memory of the simulation.
According to equation \eqref{eq:nof_basis_functions}
the number of DOF per cell for 3D DG is $1/6 p(p+1)(p+2)$. 
For two-dimensional simulations
one obtains $1/2 p(p+1)$ DOF for every cell. The total number of DOF
is thus given by $N$, $3N$, $6N$, and $10N$ for FV-2, DG-2, DG-3, and DG-4, respectively. 
At the same number of DOF, as marked by the grey dashed lines, the FV-2 and DG-2
code achieve a similar accuracy.
Moreover, when using higher order DG methods
the precision obtained per DOF is significantly increased in this test problem.

In the right-hand panel of Fig.~\ref{fig:yee_vortex}, we compare the efficiencies
of the different schemes by plotting the obtained precision as a function of the
measured run time, which directly indicates the computational
cost.\footnote{\label{foot:yee_cost} The simulations were performed
  with four {\small MPI}-tasks on a conventional desktop machine, shown is the
  wall-clock time in seconds.  A similar trend could also be observed
  for larger parallel environments; while DG is a higher order scheme,
  due to its discontinuous nature the amount of required communication
  is comparable to that of a traditional second order FV method.}  By
doing so, we try to shed light on the question of the relative speed of
the methods (or computational cost required) for a given precision. Or
alternatively, this also shows the precision attained at a fixed
computational cost.

For the following discussion, we want to remark that the isentropic
vortex is a smooth 2D test problem; slightly different conclusions may
well hold for problems involving shocks or contact discontinuities, as
well as for 3D tests. The FV-2 method consumes less time than
the DG-2 method when runs with equal resolution are compared.  On the
other hand, the error of the DG-2 scheme is smaller. As
Fig.~\ref{fig:yee_vortex} shows, when both methods are quantitatively
compared at equal precision, the computational cost is essentially the
same; hence, the overall efficiency of the two second order schemes is
very similar for this test problem. Interestingly, the $64^2$-cell FV-2
run (yellow triangle) and the $32^2$-cell DG-2 simulation (green
square) give an equally good result at almost identical runtime.

However, the higher order schemes (DG-3, DG-4) show a significant
improvement over the second order methods in terms of efficiency,
i.e. prescribed target accuracies as indicated by the grey dashed lines
can be reached much faster by
them. In particular, the third order DG (DG-3) scheme performs clearly
favourably over the second order scheme. The fourth order DG (DG-4) method
uses the SSP RK-4 time integrator, which has already five stages, in
contrast to the time integrators of the lower order methods, which
have the optimal number of $p$ stages for order $p$. Moreover, the
timestep for the higher order methods is smaller; according to
equation~\eqref{eq:time_step}, it is proportional to $1/(2p-1)$.
Nevertheless, DG-4 is the most efficient method for this
test. Consequently, for improving the calculation efficiency in smooth
regions, the increase of the order of the scheme ($p$-refinement)
should be preferred over the refinement of the underlying grid
($h$-refinement).  

For FV methods, this principle is cumbersome to achieve from a
programming point of view, because for every order a different
reconstruction scheme has to be implemented, and moreover, there are
no well-established standard approaches for implementing arbitrarily
higher order FV methods. Higher order DG methods on the other hand
can be implemented straightforwardly and in a unified way. If the
implementation is kept general, changing the order of the scheme
merely consists of changing a simple parameter. In principle, it is
also possible to do the $p$-refinement on the fly. In this way,
identified smooth regions can be integrated with large cells and high
order, whereas regions close to shocks and other discontinuities can
be resolved with many cells and a lower order integration scheme.

\subsection{Shock tube}
\label{sec:shock_tube}

Due to the non-linearity of the Euler equations, the characteristic
wave speeds of the solution depend on the solution itself.  This
dependence and the fact that the Euler equations do not diffuse
momentum can lead to an inevitable wave steepening, ultimately
producing wave breaking and the formation of mathematical
discontinuities from initially smooth states \citep[][and references
herein]{TORO}.  Such hydrodynamic shocks are omnipresent in many
hydrodynamical simulations, particularly in astrophysics. The proper
treatment of these shocks is hence a crucial component of any
numerical scheme for obtaining accurate hydrodynamical solutions.

In a standard FV method based on the RSA approach, the
solution is averaged within every cell at the end of each
timestep. The solution is then represented through piecewise constant
states that can have arbitrarily large jumps at interfaces,
consequently allowing shocks to be captured by construction.
Similarly, the numerical solution in a DG scheme is discontinuous
across cell interfaces, allowing for an accurate representation of
hydrodynamic shocks.  We demonstrate the shock-capturing abilities of
our code with a classical Sod shock tube problem \citep{SOD} in the
two-dimensional domain $(x,y)\in[0,1]^2$ with $64^2$ cells. The
initial conditions consist of a constant state on the left,
$\rho_\mathrm{l}=1$, $p_\mathrm{l}=1$, and a constant state on the
right, $\rho_\mathrm{r}=0.125$, $p_\mathrm{r}=0.1$, separated by a
discontinuity at $x=0.5$. The velocity is initially zero everywhere,
and the adiabatic index is chosen to be $\gamma=7/5$. 

In Figure~\ref{fig:shock_tube} we show the numerical and analytic
solution at $t_{\text{end}}=0.228$ for DG-3, and for
different limiting strategies.  Several problems are apparent when a
true discontinuity is approximated with higher order functions,
i.e.~the rapid convergence of the approximation at the jump is lost,
the accuracy around the discontinuity is reduced, and spurious
oscillations are introduced \citep[Gibbs phenomenon, see for example][]{ARFKEN2013}.  In the
bottom-left panel of Fig.~\ref{fig:shock_tube}, we can clearly observe
oscillations at both discontinuities, the contact and the shock.  Here
the result is obtained without any slope limiter; merely the
positivity limiter slightly adjusts higher order terms such that
negative density and pressure values in the calculation are avoided.
Nevertheless, the obtained solution is accurate at large and has even
the smallest L1 error of the three approaches tested. 

The oscillations can however be reduced by limiting the second order
terms of the solution, either expressed in the characteristic
variables (upper-left panel) or in the conserved variables (middle-left panel). 
Strikingly, the numerical solution has a considerably
higher quality when the characteristic variables are limited instead
of the conserved ones, even though in both tests the same slope-limiting 
parameters are used ($\beta=1$, $\tilde{M}=0.5$). The former
gives an overall satisfying result with a discontinuous solution
across the shock, as desired. The contact discontinuity is less sharp
and smeared out over $\mytilde 5$ cells. This effect arises from
advection errors inherent to grid codes and can hardly be
avoided. Nevertheless, the DG scheme produces less smearing compared
to an FV method once the solution is smooth, see also
Section~\ref{sec:square_advection}.  Hence, the higher order
polynomials improve the sharpness of the contact.  In the right-hand panels
of Fig.~\ref{fig:shock_tube}, we compare the mean density, pressure,
and velocity values of the numerical solution calculated with the
characteristic limiter with the analytic solutions, and find good
agreement. Especially, the shock is fitted very well by the mean values
of the computed polynomials.  To summarize, limiting of the
characteristic variables is favourable over the limiting of the
conserved variables. With the former, our DG code produces good results
in the shock tube test thanks to its higher order nature, which
clearly is an advantage also in this discontinuous problem.

\subsection{Sedov-Taylor blast wave}
\label{sec:sedov_taylor}

\begin{figure*}
\centering
\includegraphics{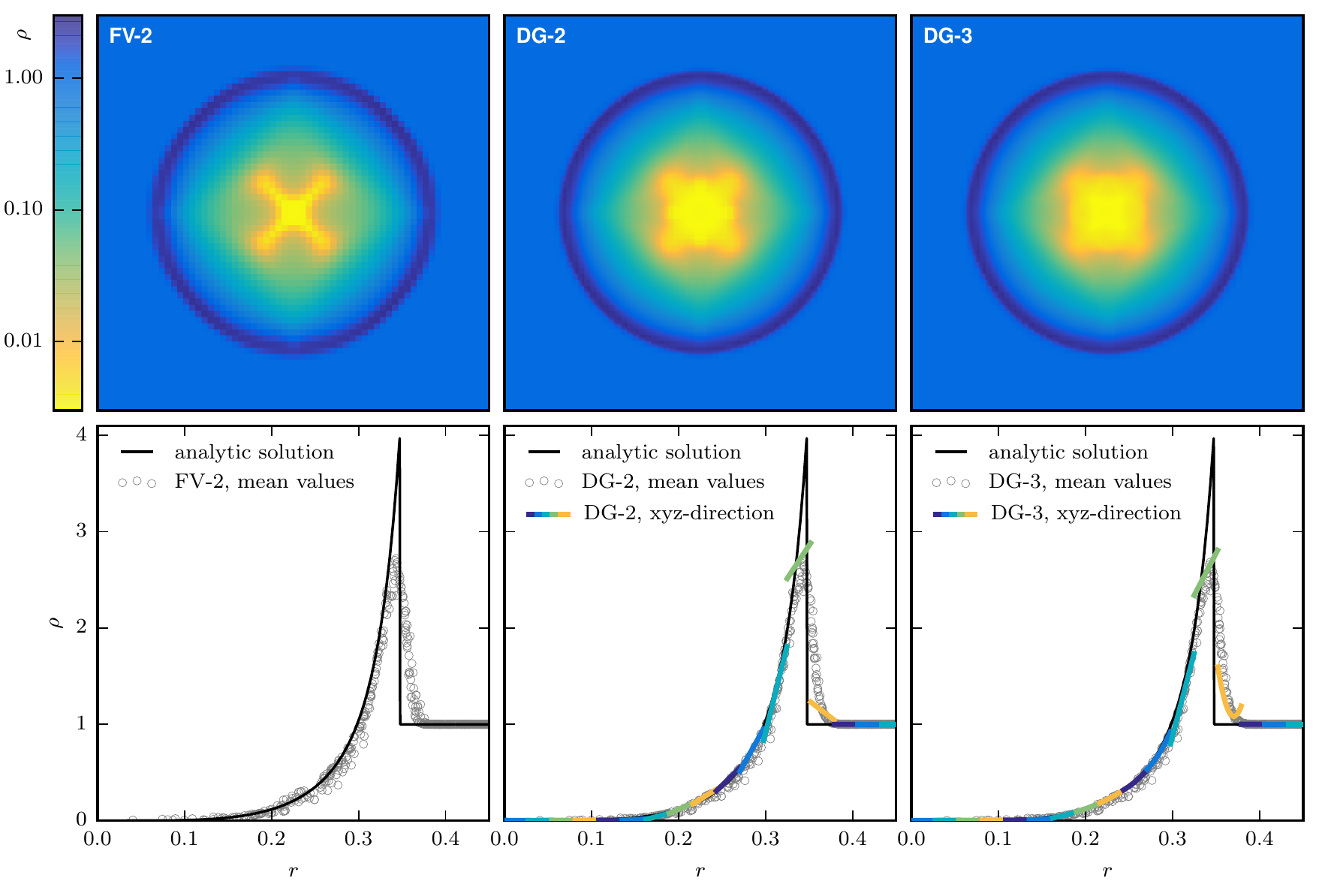}
\caption{Three-dimensional Sedov-Taylor shock wave simulations at
  $t=0.05$ calculated on a $64^3$-cell grid with FV-2 as well as with
  DG-2 and DG-3. The panels on top display central density
  slices ($z=0$) on a logarithmic scale. In this test, Cartesian grid
  codes deviate noticeably from spherical symmetry in the central low-density 
  region, and the shape of this asymmetry depends on details
  of the different methods. In the bottom panels, we compare the
  numerical results with the analytic solution. For each method, we
  plot the mean density of about every 200th cell, and for DG also
  the solution polynomials in the diagonal direction along the
  coordinates from $(0.5,0.5,0.5)$ to $(1,1,1)$.  The obtained results
  are very similar in all three methods considered here, and DG does
  not give a significant improvement over FV for this test.}
\label{fig:sedov}
\end{figure*}

\begin{figure*}
\centering
\includegraphics{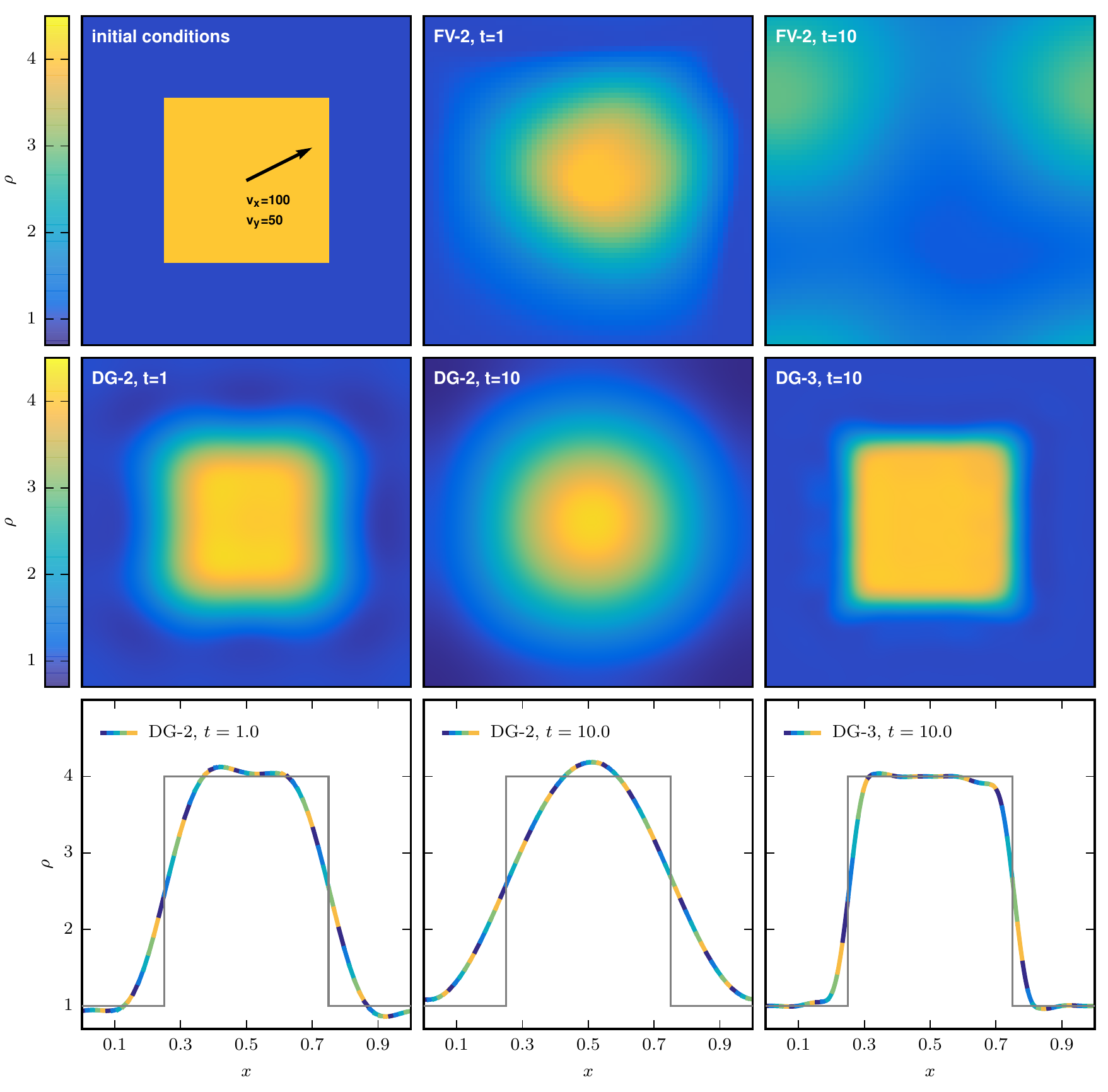}
\caption{Density maps and centred slices for the advection test with
  $64^2$ cells: a fluid with a square-shaped overdensity in
  hydrostatic equilibrium is advected supersonically, crossing the
  periodic box several hundred times.  The FV-2 method 
  shows large advection errors in this test and the square
  is smeared out completely by $t=10$.  On the other hand, the
  advection errors in the DG method become small once the solution is
  smooth. DG-3 shows less diffusion than 
  DG-2 due to the higher order representation of the advected
  shape. The time evolution of the density errors for the three
  simulations is shown in Fig.~\ref{fig:square_norms}.  }
\label{fig:square_advection}
\end{figure*}
\begin{figure}
\centering
\includegraphics{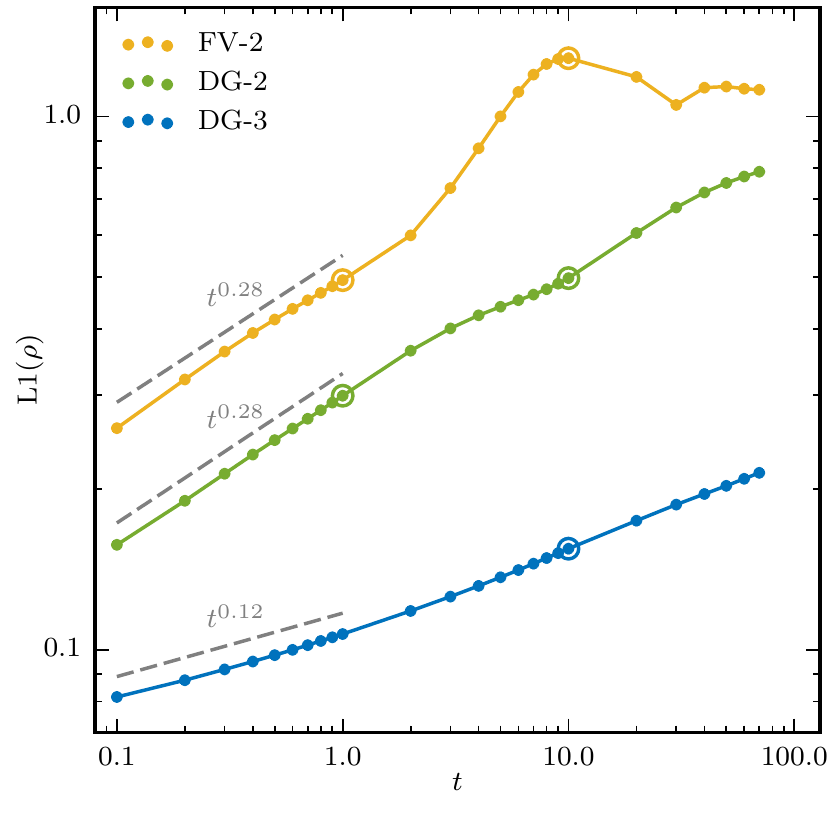}
\caption{Time evolution of the L1 density error norm for the square
  advection test. The grey lines indicate the initial slopes of error
  growth, and the encircled data points correspond to the plots shown
  in Fig.~\ref{fig:square_advection}.  Interestingly, FV-2 and DG-2 have the same
  polynomial growth, though the absolute error is smaller by a factor
  of $\approx 1.5$ for DG. On the other hand, the DG-3 method 
  does not only decrease the absolute error in this
  test, but the error growth rate is also significantly smaller.  }
\label{fig:square_norms}
\end{figure}

The previous test involved a relatively weak shock. We now confront
our DG implementation with a strong spherical blast wave by injecting
a large amount of thermal energy $E$ into a point-like region of
uniform and ultra-cold gas of density $\rho$.  The well-known analytic
solution of this problem is self-similar and the radial position of
the shock front as a function of time is given by
$R(t)=R_0({Et^2}/{\rho})^{1/5}$ \citep[see for
example][]{PADMANABHAN1}. The coefficient $R_0$ depends on the
geometry (1D, 2D, or 3D) and the adiabatic index $\gamma$; it can be
obtained by numerically integrating the total energy inside the shock
sphere.  Under the assumption of a negligible background pressure, the
shock has a formally infinite Mach number with the maximum density
compression $\rho_\text{max}/\rho=(\gamma+1)/(\gamma-1)$.

Numerically, we set up this test with $64^3$ cells in a
three-dimensional box $(x,y,z)\in[0,1]^3$ containing uniform gas with
$\rho=1$, $p=10^{-6}$, and $\gamma=5/3$.  The gas is initially at rest
and the thermal energy $E=1$ is injected into the eight central cells.
Fig.~\ref{fig:sedov} shows numerical solutions obtained with FV-2 as
well as with DG-2 and DG-3 at $t=0.05$. The density slices
in the top panels indicate very similar results; only the logarithmic
colour-coding reveals small differences between the methods in this
test.  Unlike Lagrangian particle codes, Cartesian grid codes have
preferred coordinate directions leading to asymmetries in the
Sedov-Taylor blast wave problem, especially in the inner low-density
region. FV-2 produces a characteristic cross in the centre. 
In DG the asymmetries are significantly weaker; only 
a mild cross and squared-shape trace of the initial geometry of energy injection
are visible. These effects can be further minimized by distributing the 
injected energy across more cells, for a point-like injection; however, they cannot 
be completely avoided in grid codes.

In the bottom panels of Fig.~\ref{fig:sedov}, we compare the numerical
results to the analytic solution obtained with a code provided by
\citet{KAMM2007}; in particular we have $R_0\approx 1.152$ for
$\gamma=5/3$ in 3D. Due to the finite and fixed resolution of our grid
codes, the numerical solutions do not fully reach the analytic peak
compression of $\rho_\text{max}=4$. More importantly, the DG method
does not provide a visible improvement in this particular test, and
furthermore, the results obtained with DG-2 and DG-3 are very similar. 
The reason behind these observations is the
aggressive slope limiting due to the strong shock, suppressing the
higher order polynomials in favour of avoiding over- and
undershootings of the solution. We note that a better result could of
course be achieved by refining the grid at the density jump and
thereby increasing the effective resolution. However, arguably the
most important outcome of this test is that the DG scheme copes with
an arbitrarily strong shock at least as well as a standard FV scheme,
which is reassuring given that DG's primary strength lies in the
representation of smooth parts of the solution.

\subsection{Square advection}
\label{sec:square_advection}

SPH, moving mesh, and mesh-free approaches can be implemented such that
the resulting numerical scheme is manifestly Galilean invariant,
implying that the accuracy of the numerical solution does not degrade
if a boost velocity is added to the initial conditions.  On the other
hand, numerical methods on stationary grids are in general not
manifestly Galilean invariant. Instead, they produce additional
advection errors when a bulk velocity is added, which have the
potential to significantly alter, e.g., the development of fluid
instabilities \citep{AREPO}. While the numerical solution is still
expected to converge towards the reference solution with increasing
resolution in this case \citep{ROBERTSON2010}, this comes at the price
of a higher computational cost in the form of a (substantial) increase
of the number of cells and an accompanying reduction of the timestep
size.  

We test the behaviour of the DG method when confronted with high
advection velocities by simulating the supersonic motion of a
square-shaped overdensity in hydrostatic equilibrium
\citep[following][]{GIZMO}. For the initial conditions, we choose a
$\gamma=7/5$ fluid with $\rho=1$, $p=2.5$, $v_x=100$, and $v_y=50$
everywhere, except for a squared region in the centre of the
two-dimensional periodic box, $(x,y)\in[0,1]^2$ with side lengths of
$0.5$, where the density is $\rho_{\mathrm{s}}=4$. The test is run
with a resolution of $64^2$ cells until $t=10$, corresponding to 1000
transitions of the square in the $x$-direction and 500 in the
$y$-direction. 

In Fig.~\ref{fig:square_advection}, we visually compare the results
obtained with DG-2, DG-3, and the FV-2
scheme. Already at $t=1$, the FV method has distorted the
square to a round and asymmetric shape, and at $t=10$, the numerical
solution is completely smeared out\footnote{\label{foot:sq_fv}In our
  FV-2 method, the overdensity moves slightly faster in the
  direction of advection and is clearly not centred any more at
  $t=10$. We have investigated this additional error and found that its
  occurrence depends on the choice of the slope limiter. Either way,
  the solution is completely washed out and the FV-2 method
  does not provide a satisfying result in this test.}.  In comparison, DG shows
fewer advection errors, and a better approximation of the initial shape
can be sustained for a longer time. Especially, the run with third order 
accuracy produces a satisfying result. Note that due to the high
advection speed the CFL timestep is very small, leading to around 1.7
and 3.3 million timesteps at $t=10$ with DG-2 and DG-3, respectively.

In the bottom panels of Fig.~\ref{fig:square_advection} we show
one-dimensional slices of the density solution polynomials from
$(x,y)=(0,0.5)$ to $(x,y)=(1,0.5)$, and in grey the initial conditions
as reference.  With DG-2, an immediate over- and
undershooting at the discontinuity can be observed owing to the
adopted non-TVD slope limiter ($\beta=1$, $\tilde{M}=0.5$).  More
interestingly, while in the FV-2 method the solution is only
washed out by diffusion, for DG-2 it is also steepened,
which leads to a higher maximum density at $t=10$ than present in the
initial conditions. This difference arises from the updating of the slopes
in DG instead of reconstructing them, and moreover, it vanishes if we
use TVD limiter parameters.  With the latter the slopes are reduced
aggressively, resulting in pure diffusion with a result similar to the
FV-2 method.  

We quantify the quality of each scheme in this test by measuring the
time evolution of the L1 density error norm, the result of which is
presented in Fig.~\ref{fig:square_norms}. The FV-2 and DG-2 
codes show the same initial polynomial
error growth, $\propto t^{0.28}$; however, the absolute error is much
smaller for DG, i.e.~the FV-2 error at $t=1$ is reached with
DG-2 only at a much later time, at $t=10$. With the DG-3
code (quadratic basis functions), the error grows only as
$\propto t^{0.12}$, leading to an acceptably small error even until
$t=70$, which corresponds to 23 million timesteps. Also with this
higher order approximation the solution is transformed to a smooth
numerical solution, but the mean cell values are less modified
compared to the second order code with linear basis functions.  

As demonstrated above, DG produces far smaller advection errors
compared to an FV method.  But how can this be understood,
especially since the DG scheme is also not Galilean invariant?  A
powerful tool for studying the behaviour of discretizations in
computational fluid dynamics is the modified equation analysis. Here,
the discrete equations are expanded by means of a Taylor series,
leading to a so-called modified equation which includes the target
equation to solve and additional terms. For example, the modified
equation of the first order upwind method for the scalar advection
equation is an advection-diffusion equation \citep{LEVEQUE}. The
scheme solves the advection equation to first order by construction,
but at the same time it effectively solves this advection-diffusion
equation to second order accuracy, explaining the large diffusion
errors when adopting this simple scheme, and pointing towards an
explanation for the observed diffusion in our FV-2 method. 
Such an analysis proves to be more difficult for DG;
nevertheless \citet{MOURA2014} recently accomplished a modified
equation analysis for the second order DG scheme applied to the linear
advection equation. In this case, the modified equation consists of a
physical mode and an unphysical mode moving at the wrong speed, which
is however damped very quickly. For upwind fluxes the leading error
term of the physical mode is diffusive and of third order, which is
better than naively expected for a second order method, and this is
likely one of the keys for the improved behaviour of DG we find. A
heuristic argument for the superiority of DG in this test is that
after an initial smoothing of the global numerical solution, it is not
only continuous inside every cell but also at the cell interfaces.  If
the solution is perfectly continuous across cell interfaces, the left
and right state entering the Riemann solver are identical, and the
calculated flux is always related by a simple Galilei boost to the
flux calculated in any other frame of reference. In this case, no
manifest differences in the conserved quantities in a cell due to the
flux calculation of the surface integral \eqref{eq:dg_term3} can arise
under a Galilean transformation, implying that advection errors must be
minimal. Nevertheless, some small averaging errors will arise in
practice if the current profile cannot be represented exactly at an
arbitrary grid position with the given set of cell basis functions.

\subsection{Keplerian disc}

\label{sec:keplerian_disc}
\begin{figure}
\centering
\includegraphics{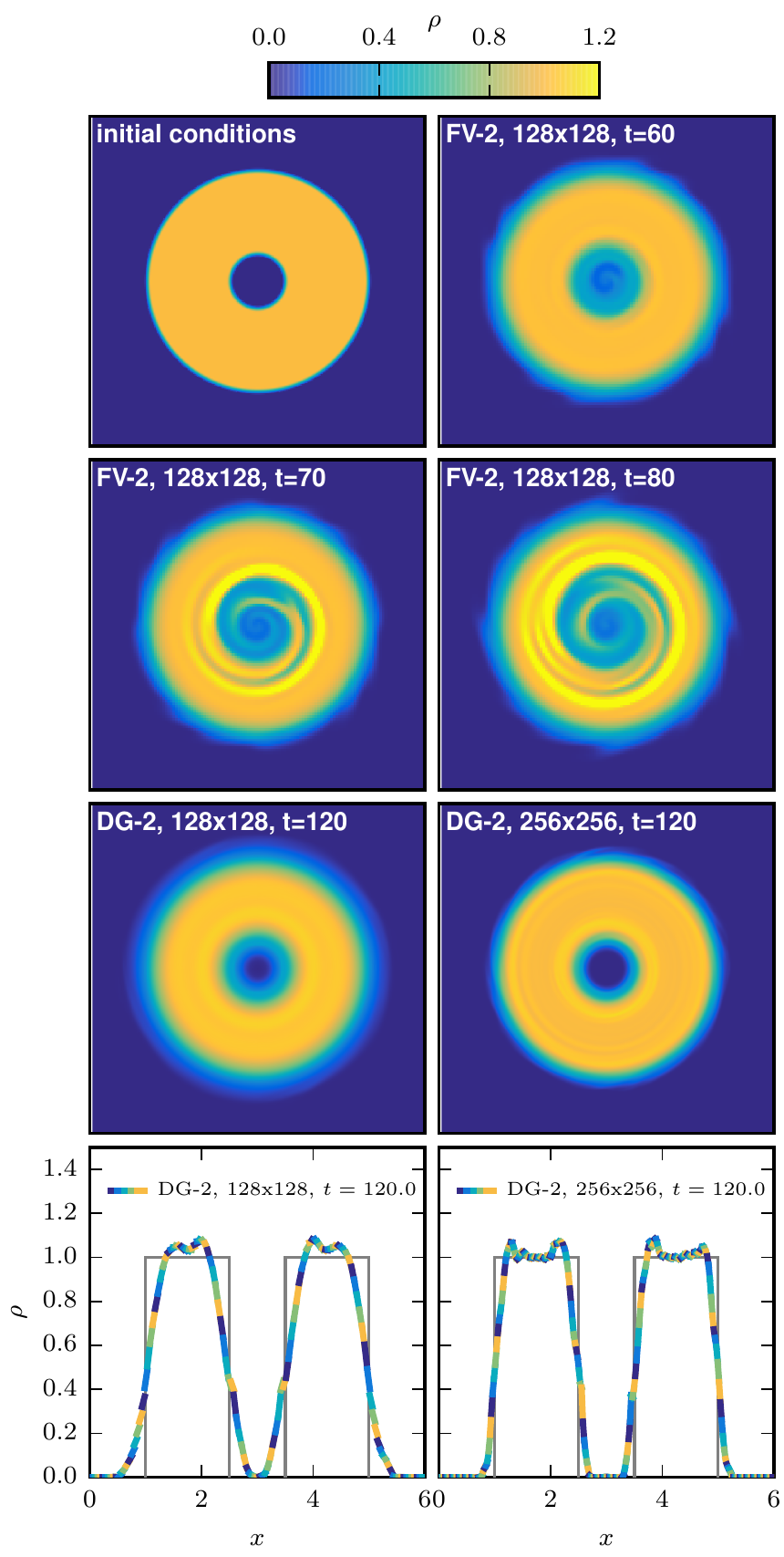}
\caption{Density evolution in the cold Keplerian disc problem. The
  centrifugal force acting on the rotating disc is balanced by an
  external static central potential such that every fluid element is
  on a Keplerian orbit.  This test is very challenging for Cartesian
  grid codes, as shear flows without pressure support are prone to
  numerical instabilities. Our FV-2 method is stable for
  around 10 orbits, at which point the disc gets disrupted eventually.
  In contrast, without a slope limiter DG conserves angular momentum
  accurately and can handle this problem very well.  }
\label{fig:kd}
\end{figure}

\begin{figure*}
\centering
\includegraphics{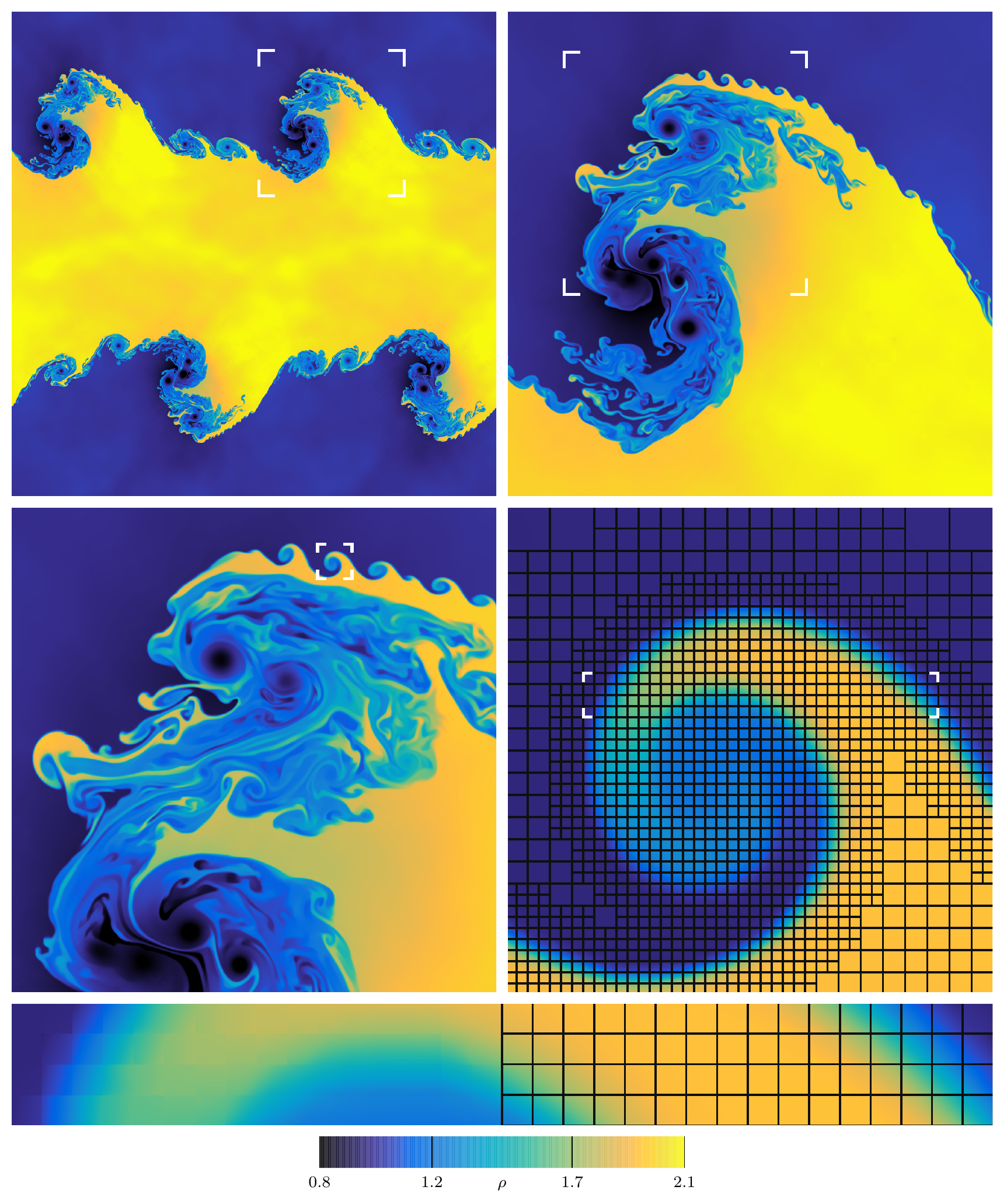}
\caption{High-resolution KH simulation with DG-4 and AMR at time $t=0.8$. The simulation
  starts with $64^2$ cells (level 6) and refines down to level 12,
  corresponding to an effective resolution of $4096^2$. We illustrate
  the AMR levels in Fig.~\ref{fig:kh_levels}.  The mesh refinement
  approach renders it possible to resolve fractal structures created
  by secondary billows on top of the large-scale waves. Furthermore,
  as can be seen in the bottom panel, the solution within every cell
  contains rich information, consisting of a third order polynomial. A
  movie of the simulation until $t=2$ may be accessed online: \url{http://youtu.be/cTRQP6DSaqA}}
\label{fig:kh}
\end{figure*}

Rotating gas discs are omnipresent in our Universe, for example in
galaxies, accretion discs around black holes, or protostellar and
protoplanetary discs.  Numerically, such objects are ideally modelled
either with a Lagrangian method or with a grid code which operates
with a suitably tailored mesh geometry and furthermore accounts for
part of the angular rotation in the solver \citep{FARGO}.  If a
simulation contains however several rotating objects at different
locations and with different orientations, as for example is the case
in cosmological galaxy formation simulations, no preferable grid
geometry exists. In this case, Cartesian grids are often adopted,
optionally with AMR, making it much more
challenging to avoid spurious errors in differentially rotating gas
flows.

These problems can be exposed by the demanding problem of a cold
Keplerian disc, where real pressure forces are negligibly small and
any spurious hydrodynamic forces from numerical errors result in
readily visible perturbations of the system.  We subject our DG code
to this test following a similar setup as recently used in other works
\citep{GIZMO, AREPO_CONVERGENCE}.  The ambient gas of the disc has
negligible density and pressure, and is initially confined to lie
between two radii. Every fluid element of the disc is rotating on a
Keplerian orbit where the centrifugal forces are balanced by an
external and static central gravitational potential. Gas self-gravity
is not included.  Analytically, the system is in perfect equilibrium
and the initial state should not change in time. The difficulty
for hydro codes applied to this test lies in the lack of pressure
support, as well as the differential rotation which leads to shearing
flows. Both can trigger numerical instabilities and the eventual
disruption of the disc. In particular, it is clear that for codes
which do not conserve angular momentum exactly, this point will
inevitably be reached eventually. In SPH codes, angular momentum is
conserved but angular momentum transport is caused by the use of
artificial viscosity, which typically is active at a small level in
strong shear flows. Using an improved switch for the viscosity,
however, the Keplerian disc problem can be integrated accurately
\citep{CULLEN2010}.

For definiteness, the initial conditions we use for our DG test are as
follows. We use the computational domain $(x,y)\in[0,6]^2$ with
periodic boundaries, and gas with an adiabatic index of
$\gamma=5/3$. The primitive variables are initially set to
\begin{align*}
p&=p_0,\\
\rho(r')&=
\begin{cases}
\rho_0&\text{if}\,\,\, r'<0.5-\frac{\Delta r}{2} \\
\frac{\rho_D-\rho_0}{\Delta r}(r'-(0.5-\frac{\Delta r}{2}))+\rho_0 &\text{if}\,\,\, 0.5-\frac{\Delta r}{2} \le r' < 0.5 + \frac{\Delta r}{2} \\
\rho_D&\text{if}\,\,\,  0.5 + \frac{\Delta r}{2} \le r' < 2 - \frac{\Delta r}{2} \\
\frac{\rho_0-\rho_D}{\Delta r}(r'-(2-\frac{\Delta r}{2}))+\rho_D&\text{if}\,\,\, 2 - \frac{\Delta r}{2} \le r' < 2 + \frac{\Delta r}{2} \\
\rho_0&\text{if}\,\,\, r' \ge 2 + \frac{\Delta r}{2}, 
\end{cases}
\end{align*}
\begin{align*}
 v_x(x',y')&=
\begin{cases}
-y'/r'^{3/2}&\text{if}\,\,\, 0.5-2\Delta r < r' < 2+2\Delta r\\
0&\text{else},
\end{cases}\\
 v_y(x',y')&=
\begin{cases}
x'/r'^{3/2}&\text{if}\,\,\, 0.5-2\Delta r < r' < 2+2\Delta r\\
0&\text{else},
\end{cases} 
\end{align*}
where the coordinates $x'$, $y'$, and $r'$ are measured in a
coordinate system with the origin at the centre $(x_0,y_0)=(3,3)$ of
the box.  The values used for the background gas are $p_0=10^{-5}$ and
$\rho_0=10^{-5}$. The disc has a density of $\rho_\mathrm{D}=1$ and a radial
extent of $[r_\text{min}=0.5,r_\text{max}=2]$, with a small transition
region of width $\Delta r=0.1$.  We adopt a time-independent external
acceleration $\myvec{a}=-\myvec{\nabla}\Phi$ with the components
\begin{align*}
a_x(x',y')&=
\begin{cases}
-x'/r'^3&\text{if}\,\,\, 0.5-0.5\Delta r< r'\\
-x'/[r'(r'^2+\epsilon^2)]&\text{else},\\
\end{cases}\\
a_y(x',y')&=
\begin{cases}
-y'/r'^3&\text{if}\,\,\, 0.5-0.5\Delta r< r'\\
-y'/[r'(r'^2+\epsilon^2)]&\text{else},\\
\end{cases}
\end{align*}
where $\epsilon=0.25$ smoothes the potential in the very inner regions
in order to avoid a singularity there. The orbital period of the
Keplerian disc depends on the radius and is given by $T=2\pi
r^{3/2}$.

We evolve the system until $t=120$, corresponding to around 20 orbits
at $r=1$, and present the results in Fig.~\ref{fig:kd}.  When the FV-2
method is used, the edges of the disc are washed out immediately but
the disc is stable for around 10 orbits. Like the majority of FV
methods in use, our scheme does not manifestly preserve angular
momentum, resulting in secular integration errors and an eventual
breakdown of the quasi-static solution. The number of orbits the disc
survives depends mainly on how carefully the problem has been set up,
as well as on the choice of slope limiter and resolution
used. However, the disruption of the disc is inevitable and can only
be delayed with adjustments of these parameters.  

On the other hand, DG schemes of second order and higher are
inherently angular momentum conserving and can hence potentially
handle this test problem much more accurately.  To test for this, we
have disabled the slope limiter of the DG scheme and use only the
positivity limiter. This is because with our simple minmod limiter,
angular momentum conservation is also violated and would result in a
similar disruption as observed with FV-2. The construction of an
improved angular momentum preserving limiting scheme is hence
desirable and worthwhile subject for future work. With the positivity
limiter alone, the DG-2 scheme can integrate the disc
stably and gives good results at $t=120$, corresponding to about 20
orbits at $r=1$. Merely two fine rings with a slightly higher density
can be observed in the inner and outer regions of the disc, which can
be attributed to the gentle overshooting of the solution at the rims
of the disc. We have also carried out this simulation with the DG-3 code.
In this case, the disc also does not break down, but
the solution shows some mild oscillations with amplitude up to 20
percent of the initial density; without applying a limiter these
cannot be suppressed.

\begin{figure}
\centering
\includegraphics{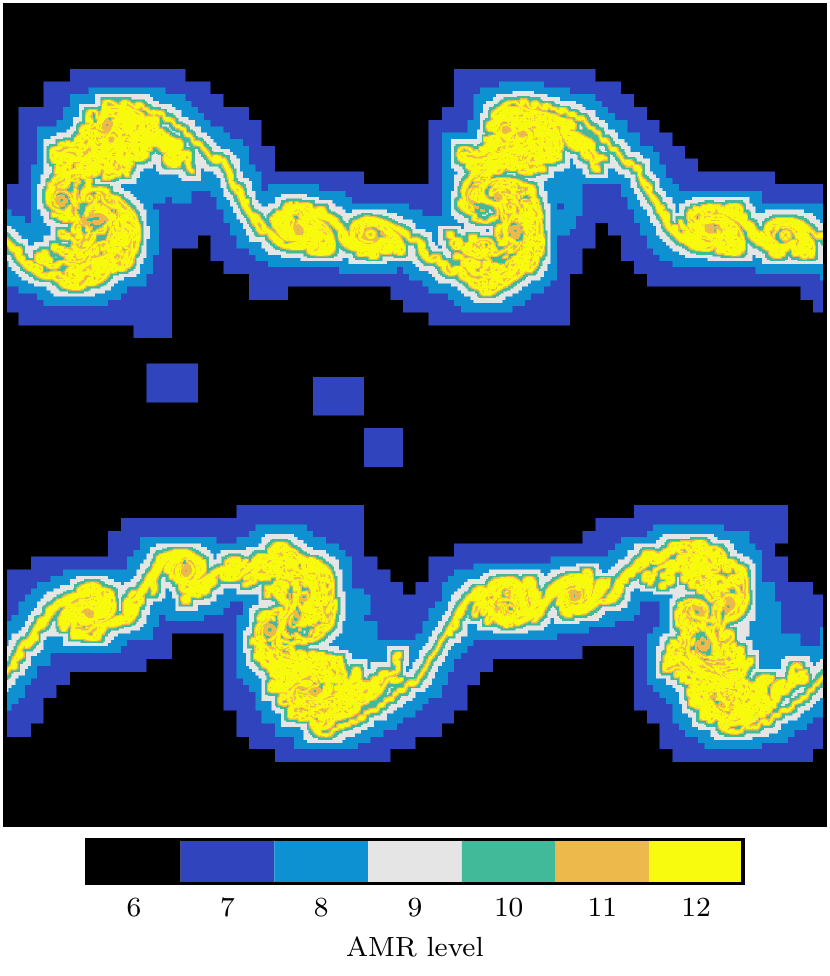}
\caption{Map of the AMR levels of the KH simulation at
  $t=0.8$. We here refine in space where the density gradient is
  steep, allowing us to capture interesting regions with high
  resolution and save computational time in places where the solution
  is smooth.}
\label{fig:kh_levels}
\end{figure}

\subsection{KH instability}
\label{sec:kh}

The KH instability is one of the most important
fluid instabilities in astrophysics, for example, it plays an
important role in the stripping of gas from galaxies falling into a
galaxy cluster.  The instability is triggered by shear flows, often
also involving fluids with different densities, and grows
exponentially until the primary billows break, subsequently leading to
a turbulent mixing of the two phases. The KH instability can be
investigated with initial conditions consisting either of a sharp
surface between two phases or with a transition region separating the
layers smoothly.  Analytic growth rates for the linear regime can be
derived for both cases \citep{CHANDRA1961, HENDRIX2014}; however, the
thickness of a smooth transition layer sets a limit on the minimum
wavelength which can become unstable in the linear phase. This
suppression is desired if one wants to compare growth rates inferred
from simulations with the analytic growth rate \citep{MCNALLY2012},
since the underlying mesh can trigger the instability of waves at the
resolution scale due to noise, a numerical effect which does not
vanish with increasing resolution.  Nevertheless, we set up a sharp
discontinuity and use the KH instability as a
high-resolution test for the robustness of our AMR implementation and
DG's capabilities of capturing small-scale turbulent structures.  

The initial conditions are chosen as in \citet{AREPO}; in the periodic
box $(x,y)\in[0,1]^2$ the primitive variables at $t=0$ are set to
\begin{align*}
p&=2.5, \\
\rho(x,y)&=
\begin{cases}
2&\text{if}\,\,\, 0.25<y<0.75 \\
1&\text{else}, \\
\end{cases} 
\end{align*}
\begin{align*}
v_x(x,y)&=
\begin{cases}
0.5&\text{if}\,\,\, 0.25<y<0.75 \\
-0.5&\text{else}, \\
\end{cases} \\
v_y(x,y)&=w_0\sin(4 \pi x) 
\left\{\exp\left[ - \frac{(y-0.25)^2}{2 \sigma^2}\right] + \exp\left[- \frac{(y-0.75)^2}{2 \sigma^2}\right]\right\},
\end{align*}
with $w_0 = 0.1$, $\sigma = 0.05/\sqrt{2}$, and the adiabatic index
$\gamma=7/5$.  The velocity perturbation in the $y-$direction supports
the development of a specific single mode on large scales. We start
with an initial resolution of $64^2$ cells (level 6) and refine where
the density slope is steep, as described in Section \ref{sec:ref_criterion}.
The maximum refinement level is set to 12, corresponding to an
effective resolution of $4096^2$ cells.  A sharp discontinuity between
the two layers in combination with AMR leads to a fast transition into
the non-linear regime and generates secondary billows early on. 

We illustrate the state of the simulation at $t=0.8$ in
Fig.~\ref{fig:kh}. The panel on the top left shows the density for
the whole two-dimensional simulation box; in the following panels we
zoom in repeatedly. The DG scheme shows only little diffusion and
mixing thanks to the higher order and the avoidance of reconstruction
steps, allowing the formation of very fine structures. Smaller
KH instabilities arise on top of the large-scale waves
demonstrating the fractal nature of this test problem. Self-similar
instabilities are present across different scales, and an ending of
this pattern is only set by the limited resolution. 

The adaptive mesh used by the calculation is overlaid in the bottom-right panel, 
revealing three different AMR levels in this subbox.  The
density gradients are well resolved with the finest AMR level (level
12), whereas smooth regions are on smaller levels.  Furthermore,
technical features of the AMR mesh structure can be seen in the plot:
the level difference between neighbouring cells is at most one, and
the mesh smoothing algorithm introduces an additional cell layer
around physically refined cells.  Fig.~\ref{fig:kh_levels} shows a
map of the AMR levels of the whole simulation box at $t=0.8$, which
can be directly compared to the top-left panel of
Fig.~\ref{fig:kh}. The number of cells at the displayed instance is
around 1.8 million cells, which corresponds to about 10 percent of the
effective level 12 resolution $4096^2$, highlighting the efficiency
gain possible with the AMR approach. Ideally, we would also like to
use a corresponding adaptiveness in time by utilizing local timesteps,
something that is planned as a code extension in future work.

\section{Summary}
\label{sec:summary}

In this work, we have developed a 3D {\small MPI}-parallel higher order
DG code with AMR for solving the Euler equations, called {\small
  TENET}, and investigated its performance in several astrophysically
relevant hydrodynamic test problems. DG methods are comparatively new
in astrophysics, despite a vast body of literature and long history of
these approaches in the applied mathematics community. A number of
highly attractive features of DG however suggest that it is timely
to introduce these methods as standard tools in computational
astrophysics. In particular, DG allows higher order to be reached
without introducing communication beyond the immediate neighbouring
cells, making it particularly well suited for parallel
computing. Also, the method is characterized by an
enhanced continuous memory access
due to the increased number of DOF per cell, making it a better match for modern computer
architectures, where floating point operations are ``almost free'' in
comparison to slow and costly repeated memory accesses of small chunks. 
In addition, DG allows an
easy and flexible way to reach arbitrarily higher order, quite unlike
FV schemes. This makes it possible to also vary the order of a scheme
in space, providing for additional flexibility in AMR approaches.

Our tests furthermore clearly show that it is computationally
worthwhile to employ higher order methods. While in general it depends
on the problem which order is optimal, we have found in our tests that
third and fourth order DG implementations are typically
computationally much more efficient compared with a second order code,
at least in regions where the solution is reasonably smooth. Moreover,
the numerical result is of higher quality also in non-smooth regions,
especially shocks are represented very well, thanks to the
discontinuous nature of the DG representation. 
They are typically captured within at most two to three cells.

Nevertheless, close to shocks and discontinuities, possible limitations of the
DG scheme become apparent. In favour of a higher overall accuracy, our DG scheme is not 
TVD, which can lead to over- and undershootings of the solution.
Moreover, higher order methods tend to produce spurious oscillations
at these locations. The detection of troubled cells and 
the prevention of this unwanted behaviour is an active topic of current research. 
The spurious oscillations in our idealized test problems 
could be prevented effectively by limiting the slopes and discarding 
higher order terms of the solution when appropriate. 
For future real-world problems, a more sophisticated method is
desirable. A very promising concept in this respect is the so-called
\mbox{$hp$-adaption}. In this approach, the grid is refined around shocks
and discontinuities, while at the same time the degree
of the polynomials is reduced, resulting in a locally more robust
scheme with less oscillations. Recently, 
\citet{SONNTAG2014} presented a practical realization of this 
concept by incorporating a 
subgrid TVD FV scheme into troubled DG cells. 
The subgrid resolution is chosen such that 
it has the same number of DOF as a DG cell.
In this way, the timesteps of the FV cells are similar
to the timesteps of the larger surrounding DG cells, and the scheme can be applied
without significant computational overhead.

The fundamental difference between DG and FV is that DG solves
directly also for higher order moments of the solution, whereas in FV
all higher order information is discarded in the implicit averaging at
the end of each timestep, necessitating a subsequent reconstruction.
This aspect of DG leads directly to two major advantages over
traditional FV methods. First, DG produces significantly less
advection errors, and furthermore, if the solution is smooth across
cell interfaces, the numerical solution does not depend on the chosen
Riemann solver. Secondly, DG inherently conserves not only mass,
momentum, and energy, but also angular momentum.  The conservation of
angular momentum is very desirable for many astrophysical
applications, e.g.~simulations involving discs, or objects like stars
or molecular clouds in rotation. There is however one caveat which has
to be kept in mind. The conservation can be spoiled by the limiting
procedure, which is reminiscent of the situation in SPH, where angular
momentum is spuriously transported by artificial viscosity. Improving
the limiter with the goal of global angular momentum conservation is
hence desirable and a promising direction for future improvements of
the DG implementation.  Finally, DG can also be comparatively easily
generalized to the AMR technique, and importantly, this can be done
without loss of accuracy, unlike in standard FV approaches. The higher
order is formally preserved due to the usage of `hanging nodes', which
are the quadrature points of interfaces between cells of different
sizes.

The
present work has focused on introducing our new code and highlighting
its differences relative to FV schemes. Future developments will focus
on more sophisticated shock capturing schemes,
scaling improvements through {\small Open-MP} hybrid parallelization,
as well as on incorporating magnetic fields and astrophysical source
terms relevant in galaxy formation. The latter is greatly facilitated
by the modular structure of {\small TENET} and its parent code {\small
  AREPO}.  In a first science application of our DG code, we 
quantitatively analyse its capabilities of capturing driven turbulence
structures (Bauer et al. 2015, in preparation).

\section*{Acknowledgements}

We thank Gero Schn\"ucke, Juan-Pablo Gallego, Johannes L\"obbert, 
Federico Marinacci, Christoph Pfrommer,
and Christian Arnold for very helpful discussions.
Furthermore, we would also like to thank the referee
for a constructive report which significantly improved this work.
We acknowledge financial support through subproject EXAMAG of
the Priority Programme 1648 `SPPEXA' of the German Science Foundation,
and through the European Research Council through ERC-StG grant
EXAGAL-308037.  KS and AB acknowledge
support by the IMPRS for Astronomy and Cosmic Physics
at the University of Heidelberg. PC was supported by the AIRBUS 
Group Corporate Foundation Chair in Mathematics of Complex 
Systems established in TIFR/ICTS, Bangalore.

\bibliography{literature} 
\bibliographystyle{mnras}

\appendix

\section{Legendre Polynomials}
\label{sec:legendre_polynomials}

We use Legendre Polynomials 
as basis functions for our discontinuous Galerkin scheme.
They can be calculated by solving Legendre's differential equation:
\begin{align}
\deriv{\xi}\left[(1-\xi^2)\deriv{\xi}P_n(\xi)\right]+n(n+1)P_n(\xi)=0, \quad n\in\mathbb{N}_0. 
\label{eq:legendre_dgl}
\end{align}
The first Legendre Polynomials are
\begin{equation}
\begin{aligned}[c]
&P_0(\xi)=1\\
&P_2(\xi)=\frac{1}{2}(3\xi^2-1)\\
&P_4(\xi)=\frac{1}{8}(35\xi^4-30\xi^2+3)\\
\end{aligned}
\begin{aligned}[c]
\,\,&P_1(\xi)=\xi\\
\,\,&P_3(\xi)=\frac{1}{2}(5\xi^3-3\xi)\\
\,\,&P_5(\xi)=\frac{1}{8}(63\xi^5-70\xi^3+15\xi).\\
\end{aligned}
\label{eq:legendre_polys}
\end{equation}
They are pairwise orthogonal
to each other, and moreover, we define the 
scaled polynomials
\begin{align}
\tilde{P}(\xi)_n=\sqrt{2n+1}P(\xi)_n
\end{align}
such that
\begin{align}
\int_{-1}^{1}\!\tilde{P}_i(\xi)\tilde{P}_j(\xi)\,\mathrm{d}\xi=
\begin{cases}
0&\text{if}\,\,\, i\ne j\\
2&\text{if}\,\,\, i=j.
\end{cases}
\end{align}

\section{Gauss-Legendre quadrature}
\label{sec:gaussian_quadrature}

The numerical integration of a function $f : [-1,+1] \rightarrow \mathbb{R}$
with the Gaussian quadrature rule of $n$ points is given by a weighted sum, 
\begin{align}
\int_{-1}^{+1}\!f(\xi)\,\mathrm{d}\xi \approx \sum_{q=1}^{n} f(\xi_q^{\oned}) \omega_q^{\oned}.
\end{align}
Here, $\xi_q^{\oned}\in(-1,+1)$ are the Gaussian quadrature nodes and $\omega_q^{\oned} $ 
are the corresponding weights. To integrate a 2D function $f : [-1,+1]^2 \rightarrow \mathbb{R}$ the tensor product
of the $n$ Gauss points can be used, viz.
\begin{align}
\nonumber&\int_{-1}^{+1}\int_{-1}^{+1}\!f(\xi_1,\xi_2)\,\mathrm{d}\xi_1\mathrm{d}\xi_2\\
&\approx \sum_{q=1}^{n}\sum_{r=1}^{n} f(\xi_{1,q}^{\oned},\xi_{2,r}^{\oned}) \omega_q^{\oned}\omega_r^{\oned}= \sum_{q=1}^{n^2}f(\myvec{\xi}_q^{\twod}) \omega_q^{\twod}.
\end{align}
The $n$-point Gaussian quadrature rule is exact for polynomials of degree up to $2n-1$,
and the one-dimensional nodes are given as the roots
of the Legendre polynomial $P_n(\xi)$. We calculate them numerically
by means of the Newton-Raphson method. As starting values of the 
iterative root finding approximate expressions for the roots can be used \citep[see for example][]{LETHER95},
\begin{align}
\xi_{q}\approx\left(1-\frac{1}{8n^2}+\frac{1}{8n^3}\right)\cos\left(\pi\frac{4q-1}{4n+2}\right),\quad q=1,\ldots,n.
\end{align}
Furthermore, the corresponding weights can be calculated as \citep{ABRAMOWITZ2012} 
\begin{align}
\omega_q=\frac{2}{(1-\xi_q^2)P_n'(\xi_q)^2},\quad q=1,\ldots,n.
\end{align}
With this approach, we can compute and store the necessary quadrature data
in the initialization routine of our DG code 
for arbitrary spatial order.

\section{Gauss-Legendre-Lobatto quadrature}
\label{sec:gll_quadrature}

Compared with Gaussian quadrature, the GLL quadrature rule is 
very similar but
includes also the endpoints of the integration interval. 
Therefore, the $n$-point GLL rule is exact for polynomials of degree $2n-3$.
The nodes are the roots $\hat{\xi}_q$ of the function $(1-\xi^2)P'_{n-1}(\xi)$,
and the corresponding weights are given by \citep{ABRAMOWITZ2012}
\begin{align}
\hat{\omega}_q=\frac{2}{n(n-1)P_{n-1}(\hat{\xi}_q)^2},\quad q=2,\ldots,n-1.
\end{align}
The weights of the endpoints are equal, $\hat{\omega}_1=\hat{\omega}_n$,
and the sum of the weights is $\sum_{q=1}^n\hat{\omega}_q=2$.

\section{Strong stability preserving Runge-Kutta methods}
\label{sec:runge_kutta}

\begin{table}
\begin{minipage}{84mm}
\begin{tabular}{ c | c c c c c}
  $c_1$ &  &  & & & \\
  $c_2$ & $a_{21}$ &  & & &\\
  $c_3$ & $a_{31}$ & $a_{32}$ & & &\\
  $\vdots$ & $\vdots$ & $\vdots$ & $\ddots$ &  &\\
  $c_s$ & $a_{s1}$ & $a_{s2}$ & $\cdots$ & $a_{s,s-1}$ &\\
  \hline
   & $b_1$ & $b_2$ & $\cdots$ & $b_{s-1}$ & $b_s$\\
\end{tabular}
\caption{Runge-Kutta butcher tableau.
\label{tab:butcher}}
\end{minipage}
\end{table}

\begin{table}
\begin{minipage}{42mm}
\begin{tabular}{ c | c c c}
\ \\
  0 &  &   \\
  1 & 1 &   \\
  \hline
   & $\frac{1}{2}$ & $\frac{1}{2}$  \\
\end{tabular}
\caption{SSP-RK2.
\label{tab:RK_2}}
\end{minipage}
\begin{minipage}{42mm}
\begin{tabular}{ c | c c c}
  0 &  &  & \\
  1 & 1 &  & \\
  $\frac{1}{2}$ & $\frac{1}{4}$ & $\frac{1}{4}$ & \\
  \hline
   & $\frac{1}{6}$ & $\frac{1}{6}$ & $\frac{2}{3}$ \\
\end{tabular}
\caption{SSP-RK3.
\label{tab:RK_3}}
\end{minipage}
\end{table}

\begin{table*}
\begin{minipage}{126mm}
\begin{tabular}{ c | c c c c c}
  0 &  &  & & & \\
  0.39175222700392 & 0.39175222700392 &  & & &\\
  0.58607968896779 & 0.21766909633821 & 0.36841059262959 & & &\\
  0.47454236302687 & 0.08269208670950 & 0.13995850206999 & 0.25189177424738 &  &\\
  0.93501063100924 & 0.06796628370320 & 0.11503469844438 & 0.20703489864929 & 0.54497475021237 &\\
  \hline
   & 0.14681187618661 & 0.24848290924556 & 0.10425883036650 & 0.27443890091960 & 0.22600748319395\\
\end{tabular}
\caption{SSP-RK4.
\label{tab:RK_4}}
\end{minipage}
\end{table*}

Let $w(t)$ be the unknown scalar solution
of the ordinary differential equation
\begin{align}
\frac{\mathrm{d}w}{\mathrm{d}t}+R(t,w)=0.
\end{align}
The propagation of the numerical solution from timestep $n$ to $n+1$
with an $s-$stage explicit RK method can be written as 
\begin{align}
w^{n+1}=w^n+\Delta t^n\sum\limits_{i=1}^sb_ik_i.
\end{align}
The factors $b_i$ weight the sum over the solution derivatives $k_i$,
which are evaluations of $R(t,w)$, viz.
\begin{align}
k_i=-R(t^n+c_i\Delta t^n, w^n+\Delta t^n(a_{i1}k_1+a_{i2}k_2+\ldots+a_{i,i-1}k_{i-1})),
\label{eq:rhs}
\end{align}
where the $c_i$ are nodes of the timestep interval.  A RK scheme is
fully specified by the weights $b_i$, the nodes $c_i$, and the RK
matrix $a_{ij}$, it can be represented in compact form by means of a
Butcher tableau (Table~\ref{tab:butcher}).  For our DG code we use
strong stability preserving RK methods, which are convex combinations
of forward Euler steps. In combination with a positivity preserving
Riemann solver and the positivity limiter outlined in
Section~\ref{sec:positivity_limiter} negative pressure and density
values can effectively be avoided in the hydro scheme.  For the second order
DG code we use SSP RK2 Heun's method (\ref{tab:RK_2}), for our
third order code the SSP RK3 from Table \ref{tab:RK_3}. These
methods are optimal in the sense that the number of stages is equal to
the order of the scheme.  It can be proven that a fourth order method
with this feature does not exist \citep{GOTTLIEB1998}, we therefore
adopt the 5-stage SSP RK 4 method tabulated in Table~\ref{tab:RK_4}
\citep{SPITERI2002}.

\section{Eigenvectors of the Euler equations}
\label{sec:eigenvectors}

The Eigenvalues of the flux Jacobian Matrix
$\myvec{\mathsf{A}}_1=\frac{\partial\myvec{f}_1(\myvec{u})}{\partial\myvec{u}}$
are $\lambda_i=\{v_1-c,v_1,v_1+c,v_1,v_1\}$. The corresponding
Eigenvectors are the columns of the matrix
\begin{align}
\mathcal{R}_x=
\begin{pmatrix}
1 & 1 & 1 & 0 & 0 \\
v_1-c & v_1 & v_1+c & 0 & 0 \\
v_2 & v_2 & v_2 & -1 & 0 \\
v_3 & v_3 & v_3 & 0 & 1 \\
h-c v_1 & k & h+c v_1 & -v_2 & v_3 \\
\end{pmatrix},
\end{align}
with the specific kinetic energy $k=\frac{1}{2}(v_1^2+v_2^2+v_3^3)$
and the specific stagnation enthalpy $h=c^2/(\gamma-1)+k$.  The left
eigenvectors of $\myvec{\mathsf{A}}_1$ are the rows of the Matrix
$\mathcal{L}_x=\mathcal{R}_x^{-1}$.  $\mathcal{L}_x$ is the linear
transformation operator from the conserved to the characteristic
variables, $\myvec{c}=\mathcal{L}_x\myvec{u}$, and can be calculated
as
\begin{align}
\mathcal{L}_x=
\begin{pmatrix}
\beta(\phi+c v_1) & -\beta(\gamma_1v_1+c) & -\beta\gamma_1v_2 & -\beta \gamma_1 v_3 & \beta \gamma_1 \\
1-2\beta\phi & 2\beta\gamma_1v_1 & 2\beta\gamma_1v_2 & 2\beta \gamma_1 v_3 & -2 \beta \gamma_1 \\
\beta(\phi-cv_1) & -\beta(\gamma_1v_1-c) & -\beta\gamma_1v_2 &  -\beta\gamma_1v_3 & \beta \gamma_1 \\
v_2 & 0 & -1 & 0 & 0 \\
-v_3 & 0 & 0 & 1 & 0 \\
\end{pmatrix},
\end{align}
where the definitions $\gamma_1=\gamma-1$, $\phi=\gamma_1 k$, and
$\beta=1/(2c^2)$ have been used.  The eigenvector matrices for the
flux Jacobians
$\frac{\partial\myvec{f}_2(\myvec{u})}{\partial\myvec{u}}$ and
$\frac{\partial\myvec{f}_3(\myvec{u})}{\partial\myvec{u}}$ are
\begin{align}
\mathcal{R}_y=
\begin{pmatrix}
1 & 1 & 1 & 0 & 0 \\
v_1 & v_1 & v_1 & 1 & 0 \\
v_2-c & v_2 & v_2+c & 0 & 0 \\
v_3 & v_3 & v_3 & 0 & -1 \\
h-c v_2 & k & h+c v_2 & v_1 & -v_3 \\
\end{pmatrix},
\end{align}
\begin{align}
\mathcal{L}_y=
\begin{pmatrix}
\beta(\phi+c v_2) & -\beta\gamma_1v_1& -\beta(\gamma_1v_2+c) & -\beta \gamma_1 v_3 & \beta \gamma_1 \\
1-2\beta\phi & 2\beta\gamma_1v_1 & 2\beta\gamma_1v_2 & 2\beta \gamma_1 v_3 & -2 \beta \gamma_1 \\
\beta(\phi-cv_2) &  -\beta\gamma_1v_1  & -\beta(\gamma_1v_2-c) &  -\beta\gamma_1v_3 & \beta \gamma_1 \\
-v_1 & 1 & 0 & 0 & 0 \\
v_3 & 0 & 0 & -1 & 0 \\
\end{pmatrix},
\end{align}
and
\begin{align}
\mathcal{R}_z=
\begin{pmatrix}
1 & 1 & 1 & 0 & 0 \\
v_1 & v_1 & v_1 & -1 & 0 \\
v_2 & v_2 & v_2 & 0 & 1 \\
v_3-c & v_3 & v_3+c & 0 & 0 \\
h-c v_3 & k & h+c v_3 & -v_1 & v_2 \\
\end{pmatrix},
\end{align}
\begin{align}
\mathcal{L}_z=
\begin{pmatrix}
\beta(\phi+c v_3) & -\beta\gamma_1v_1&  -\beta\gamma_1v_2 &-\beta(\gamma_1v_3+c) & \beta \gamma_1 \\
1-2\beta\phi & 2\beta\gamma_1v_1 & 2\beta\gamma_1v_2 & 2\beta \gamma_1 v_3 & -2 \beta \gamma_1 \\
\beta(\phi-cv_3) &  -\beta\gamma_1v_1  &   -\beta\gamma_1v_2 & -\beta(\gamma_1v_3-c) & \beta \gamma_1 \\
v_1 & -1 & 0 & 0 & 0 \\
-v_2 & 0 & 1 & 0 & 0 \\
\end{pmatrix},
\end{align}
respectively.

\section{Refinement matrices}
\label{app:refinement_matrices}

Below, we list the refinement matrices for merging the cells
$A,B,\ldots,H$ into a coarser cell
$K=\{\myvec{\xi}|\myvec{\xi}\in[-1,1]^3\}$. We calculate the integrals
by means of an exact numerical integration with $(k+1)^3$ quadrature
points before the main loop of our code.
\begin{align}
\nonumber&\text{Subcell }A=\{\myvec{\xi}|\myvec{\xi}\in[-1,0]\times[-1,0]\times[-1,0]\}:\\
&(\mathsf{P}_A)_{j,l}=\frac{1}{8}\iiint\displaylimits_{[-1,1]^3}\!\phi_j\left( \frac{\xi_1-1}{2},  \frac{\xi_2-1}{2}, \frac{\xi_3-1}{2}\right)\phi_l(\xi_1,\xi_2,\xi_3)\,\mathrm{d}\myvec{\xi}.
\end{align}
\begin{align}
\nonumber&\text{Subcell }B=\{\myvec{\xi}|\myvec{\xi}\in[0,1]\times[-1,0]\times[-1,0]\}:\\
&(\mathsf{P}_B)_{j,l}=\frac{1}{8}\iiint\displaylimits_{[-1,1]^3}\!\phi_j\left( \frac{\xi_1+1}{2},  \frac{\xi_2-1}{2}, \frac{\xi_3-1}{2}\right)\phi_l(\xi_1,\xi_2,\xi_3)\,\mathrm{d}\myvec{\xi}.
\end{align}
\begin{align}
\nonumber&\text{Subcell }C=\{\myvec{\xi}|\myvec{\xi}\in[-1,0]\times[0,1]\times[-1,0]\}:\\
&(\mathsf{P}_C)_{j,l}=\frac{1}{8}\iiint\displaylimits_{[-1,1]^3}\!\phi_j\left( \frac{\xi_1-1}{2},  \frac{\xi_2+1}{2}, \frac{\xi_3-1}{2}\right)\phi_l(\xi_1,\xi_2,\xi_3)\,\mathrm{d}\myvec{\xi}.
\end{align}
\begin{align}
\nonumber&\text{Subcell }D=\{\myvec{\xi}|\myvec{\xi}\in[0,1]\times[0,1]\times[-1,0]\}:\\
&(\mathsf{P}_D)_{j,l}=\frac{1}{8}\iiint\displaylimits_{[-1,1]^3}\!\phi_j\left( \frac{\xi_1+1}{2},  \frac{\xi_2+1}{2}, \frac{\xi_3-1}{2}\right)\phi_l(\xi_1,\xi_2,\xi_3)\,\mathrm{d}\myvec{\xi}.
\end{align}
\begin{align}
\nonumber&\text{Subcell }E=\{\myvec{\xi}|\myvec{\xi}\in[-1,0]\times[-1,0]\times[0,1]\}:\\
&(\mathsf{P}_E)_{j,l}=\frac{1}{8}\iiint\displaylimits_{[-1,1]^3}\!\phi_j\left( \frac{\xi_1-1}{2},  \frac{\xi_2-1}{2}, \frac{\xi_3+1}{2}\right)\phi_l(\xi_1,\xi_2,\xi_3)\,\mathrm{d}\myvec{\xi}.
\end{align}
\begin{align}
\nonumber&\text{Subcell }F=\{\myvec{\xi}|\myvec{\xi}\in[0,1]\times[-1,0]\times[0,1]\}:\\
&(\mathsf{P}_F)_{j,l}=\frac{1}{8}\iiint\displaylimits_{[-1,1]^3}\!\phi_j\left( \frac{\xi_1+1}{2},  \frac{\xi_2-1}{2}, \frac{\xi_3+1}{2}\right)\phi_l(\xi_1,\xi_2,\xi_3)\,\mathrm{d}\myvec{\xi}.
\end{align}
\begin{align}
\nonumber&\text{Subcell }G=\{\myvec{\xi}|\myvec{\xi}\in[-1,0]\times[0,1]\times[0,1]\}:\\
&(\mathsf{P}_G)_{j,l}=\frac{1}{8}\iiint\displaylimits_{[-1,1]^3}\!\phi_j\left( \frac{\xi_1-1}{2},  \frac{\xi_2+1}{2}, \frac{\xi_3+1}{2}\right)\phi_l(\xi_1,\xi_2,\xi_3)\,\mathrm{d}\myvec{\xi}.
\end{align}
\begin{align}
\nonumber&\text{Subcell }H=\{\myvec{\xi}|\myvec{\xi}\in[0,1]\times[0,1]\times[0,1]\}:\\
&(\mathsf{P}_H)_{j,l}=\frac{1}{8}\iiint\displaylimits_{[-1,1]^3}\!\phi_j\left( \frac{\xi_1+1}{2},  \frac{\xi_2+1}{2}, \frac{\xi_3+1}{2}\right)\phi_l(\xi_1,\xi_2,\xi_3)\,\mathrm{d}\myvec{\xi}.
\end{align}

% Don't change these lines
\bsp	% typesetting comment
\label{lastpage}
\end{document}